%% file: MFG_on_Trees_submission_APFM_publish.tex
\documentclass[a4,11pt]{article}
\usepackage{graphicx} 
\usepackage{amsmath,amsthm}
\usepackage{amsfonts}
\usepackage{latexsym}
\usepackage{amssymb}
\usepackage{setspace}
\usepackage{comment}
\usepackage{ascmac}
\usepackage{cases}
\usepackage{booktabs}
\usepackage{enumitem}
\usepackage{subcaption} 
\usepackage{float}   
\usepackage{color}
\newtheorem*{remark*}{Remark}

\makeatletter
 \renewcommand{\theequation}{
   \thesection.\arabic{equation}}
  \@addtoreset{equation}{section}
 \makeatother
\begin{document}
\title{Mean-Field Price Formation on Trees \\
{with Multi-Population and Non-Rational Agents}
\footnote{Forthcoming in {\it Asia-Pacific Financial Markets}.}
}

\author{
Masaaki Fujii\footnote{mfujii@e.u-tokyo.ac.jp, Graduate School of Economics, The University of Tokyo, Tokyo, Japan }
~\footnote{The author is not responsible in any manner for any losses caused by the use of any contents in this research.
}
}

\date{  \small
This version: July 15, 2026
}
\maketitle


\input{Fmacro-2015.tex}

\def\calf{{\cal F}}
\def\wt{\widetilde}
\def\mbb{\mathbb}
\def\ol{\overline}
\def\ul{\underline}
\def\sign{{\rm{sign}}}
\def\wh{\widehat}
\def\mg{\mathfrak}
\def\display{\displaystyle}

\def\vr{\varrho}
\def\ep{\epsilon}

\def\Prb{\mbb{P}}
\def\del{\delta}
\def\Del{\Delta}

\def\deln{\delta_{\mathfrak{n}}}
\def\oldeln{\overline{\delta}_{\mathfrak{n}}}
\def\vep{\varepsilon}
\def\bS{{\bf{S}}}
\def\bs{{\bf{s}}}

\def\red{\textcolor{red}}
\def\Ito{{It\^o}~}
\def\blangle{\bigl\langle}
\def\Blangle{\Bigl\langle}
\def\brangle{\bigr\rangle}
\def\Brangle{\Bigr\rangle}
\def\bi{\begin{itemize}}
\def\ei{\end{itemize}}
\def\ac{\acute}
\def\pr{\prime}
\def\mgn{\mathfrak{n}}
\def\mdp{\mathfrak{p}}
\def\mdq{\mathfrak{q}}

\def\part{\partial}
\def\ul{\underline}
\def\ol{\overline}
\def\vp{\varpi}
\def\nn{\nonumber}
\def\be{\begin{equation}}
\def\ee{\end{equation}}
\def\bea{\begin{eqnarray}}
\def\eea{\end{eqnarray}}
\def\bg{\boldsymbol}
\def\bull{$\bullet~$}
\def\ex{\mbb{E}}
\def\opb{\wh{\beta}}
\def\zo{{0,1}}
\def\gmone{{\gamma^1}}
\vspace{-5mm}

\begin{abstract}
This work solves the equilibrium price formation problem for the risky stock by combining mean-field game theory 
with the binomial tree framework, adapting the classic approach of Cox, Ross \& Rubinstein. For agents with exponential and recursive utilities of
 exponential-type, we prove the existence of a unique mean-field market-clearing equilibrium and derive an explicit analytic formula for equilibrium
transition probabilities of the stock price on the binomial lattice. The agents face stochastic terminal liabilities and incremental endowments that 
depend on unhedgeable common and idiosyncratic factors, in addition to the stock price path. We also incorporate an external order flow.
Furthermore, the analytic tractability of the proposed approach
allows us to extend the framework in two important directions: First, we incorporate
multi-population heterogeneity, allowing agents to differ in functional forms for their liabilities, endowments, and risk coefficients. 
Second, we relax the rational expectations hypothesis by modeling agents operating under subjective probability measures 
which induce stochastically biased views on the stock transition probabilities. 
Our numerical examples illustrate the qualitative effects of these components
on the equilibrium price distribution.
\end{abstract}
\vspace{0mm}
{\bf Keywords:}
mean-field game,  subjective measure, non-rational agents , multiple populations, recursive utility,  market-clearing

\vspace{-2mm}
\section{Introduction}
Mean-field game (MFG) theory was pioneered independently by the seminal works 
of Lasry \& Lions~\cite{Lions-1, Lions-2, Lions-3} and by those of Huang et al.~\cite{Huang-1, Huang-2, Huang-3} in the mid- to late-2000s.
These works are based on analytic methods using coupled Hamilton-Jacobi-Bellman (HJB) and Kolmogorov equations.
Subsequently, the probabilistic approach to MFG theory,  based on forward-backward stochastic differential equations (FBSDEs)
of McKean-Vlasov type,  was established by Carmona \& Delarue~\cite{Carmona-Delarue-1, Carmona-Delarue-2}.
The two volumes by the same authors, \cite{Carmona-Delarue-T1} and \cite {Carmona-Delarue-T2}, 
provide comprehensive details on the probabilistic approach and its applications.

The primary advantage of  MFG theory is its ability to transform a complex problem of stochastic differential games
among many agents into a standard optimization and fixed-point problem.
A growing body of literature attempts to solve many-agent problems
using the MFG framework. MFG theory requires, in principle, symmetric interactions among  the agents.
A particularly large number of MFG applications can be found in financial and energy markets
because symmetric interactions are standard in these settings.
There are also many economic applications of mean-field games, 
in particular, those focusing on general equilibrium models of growth, inequality and unemployment, 
dynamic demand response,  and persuasion problems among others. 
See, for example, \cite{Achdou-1,Achdou-2,  Aid-1, Aid-2, Bayraktar, Gabaix} and the references therein.

In recent years, there have also been major advances in MFG theory for applications in the equilibrium price-formation problem,
where the asset price process is endogenously constructed to ensure that supply and demand always balance among the heterogeneous
but exchangeable agents, rather than being exogenously given.
Gomes \& Sa\'ude~\cite{Gomes-Saude} present a deterministic model
of electricity price. Its extension with random supply is given by Gomes et al.~\cite{Gomes-random-supply}.
Ashrafyan et al.~\cite{Gomes-Duality} propose a duality approach transforming these problems
into variational ones that are numerically more tractable.
Shrivats et al.~\cite{Shrivats} deal with a price formation problem for
the solar renewable energy certificate (SREC) by solving FBSDEs of McKean-Vlasov type.
F\'eron et al.~\cite{Feron-Tankov} develop a tractable equilibrium model for intraday electricity markets.
Sarto et al.~\cite{Giulia} study cap-and-trade pollution regulation and derive
the equilibrium price for the carbon emission.
Regarding the price formation of securities, Fujii \& Takahashi~\cite{Fujii} show
that the equilibrium price process can be characterized by  FBSDEs of  conditional
McKean-Vlasov type. Its strong convergence to the mean-field limit from a finite-agent setting
is proved in \cite{Fujii-SC}, and its extension to the presence of a major player is given in \cite{Fujii-Major} by the same authors. 
Fujii~\cite{Fujii-CP} develops a model that allows the co-presence of cooperative and non-cooperative populations to
investigate how the price process is formed when the agents in one population act in a coordinated manner.

There remain two mathematical limitations in the above results: firstly, the
relevant control of each agent is the trading rate and hence their asset position is constrained to follow an absolutely continuous process with
respect to the Lebesgue measure $dt$; secondly, the objective function of each agent consists of penalties on the 
trading speed and on the inventory size of the assets. Therefore, the above 
frameworks cannot deal with the general self-financing trading strategies nor with the utility functions defined directly in terms of  the 
wealth process of the portfolio. A major obstacle in dealing with a utility function of the wealth resides in the difficulty of guaranteeing
the convexity of the Hamiltonian associated with the Pontryagin’s maximum principle and in the difficulty of  getting 
enough regularity to prove the well-posedness of the associated FBSDEs.
These problems are solved by Fujii \& Sekine~\cite{Fujii-Sekine1, Fujii-Sekine2} for the agents with exponential utilities by applying  the method of Hu, Imkeller \& M\"uller~\cite{Hu-Imkeller}
based on  the  martingale optimality principle. An extension to the setting with partial information is studied by Sekine~\cite{Sekine}.
They find  that a novel quadratic-growth backward SDE (qg-BSDE) of conditional McKean-Vlasov type characterizes the equilibrium risk-premium process. Unfortunately, however, the existence of the solution of this mean-field qg-BSDE has been proved
only under rather restrictive conditions. This is because the conditional McKean-Vlasov nature of the equation
makes the classical approach of  Kobylanski~\cite{Kobylanski} no longer applicable.
Moreover, even if the well-posedness of the equation were to be completely solved, its
numerical evaluation would remain prohibitively difficult. This is a major hurdle
for practical applications, a limitation shared by all existing works of continuous-time framework.

In this work, we study the price-formation problem for a single risky stock. To understand how the equilibrium price 
process changes its behavior due to the market environment and the differences in the distribution of characteristics 
among the agents, it is necessary to obtain more explicit and numerically tractable solutions than those in the existing 
literature. To this end, we propose a framework that combines MFG theory with the classical idea of binomial trees,
initiated by Sharpe~\cite{Sharpe} and formalized in Cox, Ross \& Rubinstein~\cite{Cox}.
By leveraging the discrete structure of a binomial lattice, we can explicitly determine the equilibrium transition probabilities
that clear the stock market in a straightforward way. This discrete-time, tree-based approach 
is the key technical device that allows us to incorporate more general market environments 
without sacrificing the tractability of the equilibrium solution.

The agents in our model---trading desks of financial and investment firms---have standard exponential utilities or recursive utilities 
of the exponential type. We allow for general stochastic terminal liabilities and incremental endowments that depend on 
unhedgeable common factors $Y$  and idiosyncratic factors $Z^i$, as well as the stock price path.
The incremental endowments represent non-tradable income from a firm's other business activities. We also 
analyze the impact of external order flow from other market participants or a major financial player.
The presence of unhedgeable risk factors $(Y,Z^i)$ in these financial items
creates an incomplete market for the agents, 
motivating their demand for the risky stock to hedge their {\it coupled} exposures.

The explicit solutions obtained through our binomial tree approach provide significant analytical and numerical tractability, 
allowing us to extend the framework in two important ways that would be a formidable challenge in the continuous-time setting:
\begin{itemize}[noitemsep]
\item Addressing Market Heterogeneity: We readily extend the model to a multi-population equilibrium ($\text{Section}~\ref{sec-multi-p}$), 
capturing the structural diversity of the financial sector by allowing different populations to have distinct functional forms for liabilities, endowments, and risk characteristics.
\item Relaxing Rational Expectations: We generalize the model to agents operating under subjective probability measures ($\text{Section}~\ref{sec-subjective}$), which induce stochastically biased state-dependent views on the stock price transition probabilities. 
This extension is in the spirit of the recent development in $\text{MFG}$ theory initiated by Moll \& Ryzhik~\cite{Moll-irrational}, 
and relaxes the standard rational expectations hypothesis implicitly adopted in most of the $\text{MFG}$ literature.
\vspace{-0.2em}
\end{itemize}
Our explicit solutions for the equilibrium transition probabilities enable us to numerically evaluate the marginal and conditional equilibrium price distributions with respect to the common macroeconomic factors $Y$. Our numerical examples reveal 
the qualitative dependence of the equilibrium distributions on the agents' characteristics and the above-introduced elements.

The analytic tractability achieved via our binomial tree framework is the key contribution 
that overcomes the theoretical and numerical difficulties in the continuous-time approaches, 
which involve either a (coupled systems of) complex McKean-Vlasov $\text{FBSDEs}$ (or qg-BSDEs), 
or a coupled system of HJB and Kolmogorov equations.
Beyond theoretical advancement, our methodology offers a readily implementable framework for practical applications.
It provides the equilibrium excess return required to compensate agents for the risk in the stock position, 
necessary to clear the market. The ability to obtain such explicit solutions offers a valuable tool for 
regulatory bodies performing market stability analysis under various agent-based scenarios. 

We structure the rest of the paper as follows: Section~\ref{sec-terminal} investigates the mean-field equilibrium 
among the agents with exponential utilities subject to terminal liabilities. Section~\ref{sec-recursive} deals with an extension 
to recursive utilities and also introduces cash spending (i.e., nominal consumption) and incremental endowments.
Section~\ref{sec-multi-p} studies an extension to the presence of multiple populations.
Section~\ref{sec-subjective} extends the previous frameworks to incorporate agents operating under respective subjective measures
which induce state-dependent stochastic biases.
Section~\ref{sec-numerical} provides several illustrative numerical examples and examines their important implications.
Section~\ref{sec-conclude} summarizes our findings and discusses possible directions for future research.
\section{Exponential utility with terminal liability}
\label{sec-terminal}
We begin our investigation into the mean-field price formation problem by considering a simple model with a countably infinite number of agents
possessing exponential utility. These agents are optimizing their wealth with terminal liability
by carrying out self-financing trading  on a deterministic money market account
and a single risky stock.
Each agent must manage financial risk arising from the common market shocks as well as their own idiosyncratic shocks.
Notably, our model incorporates non-tradable macroeconomic and/or environmental shocks that affect all the agents,
in addition to the shocks from the stock price process.

\begin{remark}[On the stock]
In this work, the financial institution's overall equity exposure is simplified to a single asset
which serves as a proxy for a broad, diversified market index (e.g., S\&P 500) to keep the focus on the mean-field price formation problem.
\end{remark}

\subsection{The setup and notation}
\label{sec-setup}
Let us start from explaining the relevant probability spaces.
 $(\Omega^0,\calf^0, (\calf_{t_n}^0)_{n=1}^N, \mbb{P}^0)$ is a complete filtered probability space, where $0=t_0<t_1<\cdots<t_N=T$
is an equally spaced time sequence using a time step $\Del:=T/N$ where $T<\infty$ and $N\in \mbb{N}$ are given constants. 
The filtration $(\calf_{t_n}^0)_{n=0}^N$ is generated by two stochastic processes, one is a strictly positive process $(S_n:=S(t_n))_{n=0}^N$
and the other is a $d_Y$-dimensional process $(Y_n:=Y(t_n))_{n=0}^N$, i.e. $\calf_{t_n}^0:=\sigma\{S_k, Y_k, 0\leq k\leq n\}$. 
In the model below, we shall use $S_n$ to denote the stock price at $t_n$ and $Y_n$ the common shocks affecting all the agents at $t_n$.
$S_0>0$ and $Y_0\in \mbb{R}^{d_Y}$ are given constants and thus $\calf_0^0$ is trivial.

In addition to the above space, 
we consider a countably infinite number  of complete filtered probability spaces $(\Omega^i, \calf^i, (\calf_{t_n}^i)_{n=0}^N,\mbb{P}^i), ~i\in \mbb{N}$.
For each $i$, $(\Omega^i,\calf^i, (\calf_{t_n}^i)_{n=0}^N,\mbb{P}^i)$ is endowed with 
$\calf^{i}_0$-measurable random variables $(\xi_i, \gamma_i)$ as well as an $(\calf^i_{t_n})_{n=0}^N$-adapted 
stochastic process $(Z_n^i=Z^i(t_n))_{n=0}^N$.  Here, $\xi_i, \gamma_i$ are both $\mbb{R}$-valued 
and $\xi_i$ is used to represent the initial wealth and $\gamma_i$ the size of absolute risk aversion
of agent-$i$. The $d_Z$-dimensional process $(Z_n^i)_{n=0}^N$ is used to model idiosyncratic shocks to each agent.
The fact that $(\xi_i, \gamma_i)$ are $\calf_0^i$-measurable means that the  agent-$i$ knows their 
initial wealth and the size of risk aversion at time zero.

By the standard procedures (see, for example, Klenke~\cite[Chapter 14]{Klenke}), the complete filtered probability space 
$(\Omega,\calf,(\calf_{t_n})_{n=0}^N,\mbb{P})$ is defined as the product of all the above spaces
\be
(\Omega,\calf,(\calf_{t_n})_{n=0}^N,\mbb{P}):=(\Omega^0,\calf^0,(\calf_{t_n}^0)_{n=0}^N,\mbb{P}^0)
\otimes_{i=1}^\infty (\Omega^i,\calf^i,(\calf_{t_n}^i)_{n=0}^N,\mbb{P}^i) \nn
\ee  
which denotes the full probability space containing the entire environment of our model.
Therefore, by construction, $((S_n),(Y_n))$ and $(\xi_i, \gamma_i, (Z_n^i)),~i\in \mbb{N}$ are mutually independent.
On the other hand, the relevant probability space for each agent-$i$ is the product probability space defined by
\be
(\Omega^{0,i},\calf^{0,i}, (\calf_{t_n}^{0,i})_{n=0}^N,\mbb{P}^{0,i}):=(\Omega^0,\calf^0,(\calf_{t_n}^0)_{n=0}^N,\mbb{P}^0)
\otimes (\Omega^i,\calf^i,(\calf_{t_n}^i)_{n=0}^N,\mbb{P}^i), \nn
\ee 
which reflects our assumption that common shocks are public knowledge, but the idiosyncratic shocks are private to 
each agent.
We shall use the same symbols,  such as $(S_n, Y_n, \gamma_i, \cdots)$, if they are 
defined as trivial extensions on larger product probability spaces. 
Expectations with respect to $\mbb{P}^0$, $\mbb{P}^i$, $\mbb{P}^{0,i}$ and $\mbb{P}$ are denoted by $\ex^0[\cdot]$, $\ex^i[\cdot]$, 
$\ex^{0,i}[\cdot]$ and $\ex[\cdot]$, respectively. 

\bigskip
In this work, we restrict the trajectories of $(S_n)_{n=0}^N$ onto a recombining binomial tree. For each $n=1, \ldots, N$,
the random variable $\wt{R}_n:=S_n/S_{n-1}$ takes only the two possible values, either $\wt{u}$ or $\wt{d}$.
This means that the set of all possible values taken by $S_n$ is given by $\cals_n:=\{S_0 \wt{u}^k\wt{d}^{n-k}, 0\leq k\leq n\}$,
which is a finite subset of $(0,\infty)$.
Let $\cals^n:=\{(s_k)_{k=0}^n\in \mbb{R}^{n+1}:\mbb{P}^0(S_k=s_k, 0\leq k\leq n)>0\}$
be the set of all values taken by the stock price trajectory $(S_k)_{k=0}^n$.
Moreover, to avoid technical issues with the conditional probabilities, 
we also assume that the process $Y$ takes values in a finite set at every $t_n$.
We use $\caly_n:=\{y\in \mbb{R}^{d_Y}: \mbb{P}^0(Y_n=y)>0\}$ to denote the set of all values taken by $Y_n$. 
We also use $\caly^n:=\{(y_k)_{k=0}^n\in \mbb{R}^{d_Y\times (n+1)}: \mbb{P}^0(Y_k=y_k,0\leq k\leq n)>0\}$ for each $0\leq n\leq N$,
to represent the set of all  values taken by the trajectory  $(Y_k)_{k=0}^n$.
In a similar manner,  we denote the range of the random variable $Z_n^i$ by $\calz_n$.
There is no need to impose the restriction of finite state space on variables other than $(S,Y)$.
The time-$t_n$ value of the risk-free money market account is given by $\exp\bigl(r n\Del\bigr) $
where $r\geq 0$ is a non-negative constant denoting the risk-free (nominal) interest rate. We also use the symbol $\beta:=\exp(r\Del)$.
For later use, let us also define
\be
u:=\wt{u}-\exp(r\Del), \quad d:=\wt{d}-\exp(r\Del) \nn
\ee
and 
\be 
R_n:=\wt{R}_n-\exp(r\Del), ~1\leq n\leq N. \nn
\ee 
The random variable $R_n$ takes the values either $u$ or $d$. 
In the following, to lighten the notational burden, we use $\ex^{0,i}\bigl[ \cdot |s,y,z^i, \gamma_i\bigr]$ to denote
$\ex^{0,i}\bigl[ \cdot |S_{n-1}=s, Y_{n-1}=y, Z_{n-1}^i=z^i, \gamma_i=\gamma_i\bigr]$ for $(s,y,z^i)\in \cals_{n-1}\times \caly_{n-1}\times\calz_{n-1}$.
With a slight abuse of notation, we shall use the same symbols for the realizations of $\calf^i_0$-measurable random variables 
($\gamma_i$ in the above example).

\begin{assumption}
\label{assumption-T1}
{\rm (i):} $\wt{u}$ and $\wt{d}$ are real constants satisfying
$0<\wt{d}<\exp(r\Del)<\wt{u}<\infty$. \\
{\rm (ii):} The variables $(\xi_i,\gamma_i, (Z_n^i)_{n=0}^N)$ are identically distributed across all agents $i=1,2,\ldots$.\\
{\rm (iii):} There exist constants $\ul{\xi}, \ol{\xi}$ and $\ul{\gamma}, \ol{\gamma}$ so that for every $i\in \mbb{N}$,
\be
\xi_i\in [\ul{\xi},\ol{\xi}]\subset \mbb{R}, \quad \gamma_i\in \Gamma:=[\ul{\gamma},\ol{\gamma}]\subset (0,\infty). \nn
\ee 
{\rm (iv):} For each $i$, $(Z_n^i)_{n=0}^N$ is a Markov process i.e., $\mbb{E}^i[f(Z_n^i)|\calf_{t_k}^i]=\ex^i[f(Z_n^i)|Z_k^i]$ 
for every bounded measurable function $f$  on $\calz_n$ and $k\leq n$.\\
{\rm (v):} $(Y_n)_{n=0}^N$ is a Markov process i.e., $\mbb{E}^0[f(Y_n)|\calf_{t_k}^0]=\ex^0[f(Y_n)|Y_k]$
for every bounded measurable function $f$ on $\caly_n$ and $k\leq n$. \\
{\rm (vi):} The transition probabilities of $(S_n)_{n=0}^N$ satisfy, for every $0\leq n\leq N-1$, a.s.,
\be
\begin{split}
\mbb{P}^0(S_{n+1}=\wt{u}S_{n}|\calf_{t_n}^0)&=\mbb{P}^0(S_{n+1}=\wt{u}S_n|S_n,Y_n)=:p_{n}(S_{n},Y_{n}), \\
\mbb{P}^0(S_{n+1}=\wt{d}S_{n}|\calf_{t_n}^0)&=\mbb{P}^0(S_{n+1}=\wt{d}S_n|S_n,Y_n)=:q_{n}(S_{n},Y_{n}), \nn
\end{split}
\ee
where $p_{n}, q_{n}~(:=1-p_n) :\cals_{n}\times \caly_n\rightarrow \mbb{R}, ~0\leq n\leq N-1$ are bounded measurable functions
satisfying 
\be
0<p_n(s,y), q_n(s,y)<1 \nn
\ee
for every $(s,y)\in \cals_n\times \caly_n$.
\end{assumption}

Let us give some remarks on the above assumptions.
Firstly, by the condition (i), we have $d<0<u$.
It is well-known that the transition probabilities under the risk-neutral measure $\mbb{Q}$
for the classical binomial framework are given by $p^{\mbb{Q}}:=(-d)/(u-d)=(\exp(r\Del)-\wt{d})/(\wt{u}-\wt{d})$ 
for the up-move and $q^{\mbb{Q}}:=1-p^{\mbb{Q}}$ for the down-move.
These probabilities are uniquely determined by the parameters $(\wt{u},\wt{d})$ and the risk-free interest rate.
In this paper, we fix the relative jump size $(\wt{u}, \wt{d})$ to be constant across all nodes; however this is done merely for simplicity.
The entire discussion of our paper still holds even if $(\wt{u},\wt{d})$ varies from node to node.
Moreover, thanks to the famous result by Derman \& Kani (1994)~\cite{Derman-Kani}, one can construct so-called ``{\it implied binomial trees}''
by adjusting the position $(\wt{u},\wt{d})$ node by node to reproduce  implied volatility surface 
in the option market,  while keeping the recombining property of binomial trees intact.
Therefore, if necessary, our discussion below can be applied to a binomial tree whose risk-neutral distribution is consistent with the option market.
The boundedness assumption in  (iii) is not crucial. One can relax it by adding appropriate integrability conditions.

Our goal in this section is to find a set of transition probabilities of the form $\bigl(p_n(s,y))_{n=0}^{N-1}$ so that the demand and 
supply of the stock are balanced among the agents at every node $(s,y)\in \cals_n\times \caly_n, 0\leq n\leq N-1$.
Note that we can assume, without any loss of generality, that the process $(Y_n)_{n=0}^N$ and $(Z_n^i)_{n=0}^N$
are Markov, since,  if necessary, we can recover Markovian property by lifting $Y, Z^i$ to higher dimensional processes.
However, the condition (vi) is not trivial. In fact, in the next section, we shall study more general
situations where the transition probability must be dependent on the past history of the stock price to achieve the market-clearing equilibrium.
Under the current condition (v) and (vi), $(S_{n+1}, Y_{n+1})$ satisfy the property:
\be
\label{Y-conditional-1}
\ex^0[f(S_{n+1})g(Y_{n+1})|\calf_{t_n}^0]=\ex^0[f(S_{n+1})|S_n,Y_n]\ex^0[g(Y_{n+1})|Y_n]~{\rm a.s.,}\quad 0\leq n\leq N-1, 
\ee
for any bounded measurable functions $f:\cals_{n+1}\rightarrow \mbb{R}$ and $g:\caly_{n+1}\rightarrow \mbb{R}$.
We can interpret that the process $(Y_n)_{n=0}^N$ represents some standalone macroeconomic and/or environmental factors 
which are not influenced by the agents' trading activities. It may naturally serve as a state process in regime switching models.

\begin{remark}
Note that the bound for the transition probabilities in (vi) guarantees the equivalence of  probability measures $\mbb{P}^0\circ S^{-1}$ and
$\mbb{Q}\circ S^{-1}$. Hence, our system is arbitrage free. 
\end{remark}

\subsection{The individual optimization problem}
To obtain the market-clearing equilibrium of the risky stock among a large number of strategically interacting agents, 
we adopt the MFG framework. 
The standard MFG approach involves two steps: (i) we assume the stock price process (or transition probabilities in our case), which is formed by the collective actions of the agents and adapted to the public information $(\calf^0_{t_n})_{n=0}^N$, is given
and then solve each agent's optimization problem as a price taker; (ii) we solve the fixed-point problem of the consistency
condition, (i.e., the market-clearing), to obtain the equilibrium price distribution. 
In this subsection, we address the optimization task of step (i).

We now formulate the optimization problem for each agent. Agent-$i$, for $i\in\mbb{N}$,  with an initial wealth $\xi_i$, 
engages in self-financing trading involving the risk-free money market account and a single risky stock.
They adopt an $(\calf_{t_n}^{0,i})_{n=0}^{N-1}$-adapted trading strategy $(\phi_n^i)_{n=0}^{N-1}$, representing the 
invested amount of cash in the stock at time $t_n$. The associated wealth process of agent-$i$, denoted by $(X_n^i=X^i(t_n))_{n=0}^N$,
follows the dynamics
\be
\begin{split}
X_{n+1}^i&=\exp(r\Del)(X_n^i-\phi_n^i)+\phi_n^i \wt{R}_{n+1}=\beta X_n^i+\phi_n^i R_{n+1}, \nn
\end{split}
\ee
where $X_0^i=\xi_i$ and $\beta=\exp(r\Del)$. Recall that $R_{n+1}:=\wt{R}_{n+1}-\exp(r\Del)$.

Each agent-$i$ is supposed to solve the optimization problem:
\be
\label{problem-terminal}
\begin{split}
\sup_{(\phi_n^i)_{n=0}^{N-1}\in \mbb{A}^i}\ex^{0,i}\Bigl[-\exp\Bigl(-\gamma_i \bigl(X_N^i-F(S_N,Y_N,Z_N^i)\bigr)\Bigr)|\calf_0^{0,i}\Bigr], 
\end{split}
\ee
where 
$
\mbb{A}^i:=\bigl\{(\phi_n^i)_{n=0}^{N-1}: \phi_n^i \text{ is an $\calf_{t_n}^{0,i}$-measurable real random variable} \bigr\} \nn
$
denotes the admissible control space.  
Here, we assume that agent-$i$ has full knowledge of the common market information and their own private idiosyncratic information, 
but no knowledge of the private idiosyncratic information of the other agents.

\begin{assumption}
\label{assumption-T2}
 {\rm (i):} $F:\cals_N\times \caly_N\times \calz_N \rightarrow \mbb{R}$ is a bounded measurable function. \\
{\rm (ii):} Every agent is a price-taker in the sense that they consider the stock price process (and hence its transition 
probabilities specified in Assumption~\ref{assumption-T1} (vi)) to be exogenously given  by the collective actions 
of the others and unaffected by the agent's own trading strategies.
\end{assumption}
\noindent
$F(S_N,Y_N,Z_N^i)$ denotes the stochastic terminal liability (or the negative of the terminal endowment),
depending on $S_N,Y_N$ and $Z_N^i$. 
Given  the exponential utility assumption,  the constant shift $F\rightarrow F+c$ 
does not alter the optimization problem; hence, only the dependence on $(S_N, Y_N, Z_N^i)$ 
in the function $F$ is relevant.
The condition (ii) is a plausible assumption since every agent naturally knows their individual trading share is negligible in the market.
This negligibility of individual actions is a key assumption for  the standard MFG technique,  as previously remarked.

\begin{remark}
\label{rem-F}
Let us briefly discuss the economic motivation for including the stochastic terminal liability
$F(S_N,Y_N,Z_N^i)$.  Since we primarily model trading desks of various financial firms as our agents, 
it is natural to suppose that they are subject to stochastic liabilities (such as a portfolio of derivative contracts) 
dependent on the stock price. These liabilities typically arise from customer relationships of financial firms,
and are not directly controllable by the trading desks. It is also plausible that the size of the liability varies across 
agents based on their idiosyncratic factors $(Z_N^i)$ as well as common macroeconomic/environmental factors $(Y_N)$. 
This structure applies naturally to non-financial firms as well.
\end{remark}

We now characterize the optimal trading strategy for each agent. 
Applying the well-known scheme of backward induction for discrete-time models,
but now in the presence of common as well as idiosyncratic shocks, we establish the following result.
\begin{theorem}
\label{th-t1}
Let Assumptions~\ref{assumption-T1} and \ref{assumption-T2} be in force.
Then the problem $(\ref{problem-terminal})$ has an a.s.\! unique optimal solution $(\phi^{i,*}_{n-1})_{n=1}^{N}$,
which is an adapted process defined by a bounded
measurable function $\phi_{n-1}^{i,*}:\cals_{n-1}\times\caly_{n-1}\times \calz_{n-1}\times \Gamma\rightarrow \mbb{R}$
such that $\phi_{n-1}^{i,*}:=\phi_{n-1}^{i,*}(S_{n-1},Y_{n-1},Z_{n-1}^i,\gamma_i)$ a.s.\!, where
\be
\label{th-t1-eq1}
\begin{split}
&\phi_{n-1}^{i,*}(s,y,z^i,\gamma_i):=\frac{1}{\gamma_i (u-d)}\frac{\beta^n}{\beta^N}\Bigl\{\log\Bigl(-\frac{p_{n-1}(s,y)u}{q_{n-1}(s,y)d}\Bigr)
+\log\bigl(f_{n-1}(s,y,z^i,\gamma_i)\bigr)\Bigr\}, \\
&f_{n-1}(s,y,z^i,\gamma_i):=\frac{\ex^{0,i}[V_n(s\wt{u},Y_n,Z_n^i,\gamma_i)|y,z^i,\gamma_i]}{\ex^{0,i}[V_n(s\wt{d},Y_n,Z_n^i,\gamma_i)|y,z^i,\gamma_i]}.
\end{split}
\ee
For every $1\leq n\leq N$, $f_{n-1}:\cals_{n-1}\times \caly_{n-1}\times\calz_{n-1}\times \Gamma\rightarrow \mbb{R}$ 
and $V_n:\cals_n\times \caly_n\times \calz_n\times \Gamma \rightarrow \mbb{R}$ are
measurable functions satisfying the uniform bounds
$0<c_n\leq  f_{n-1}, V_n \leq C_n<\infty$ on their respective domains,  for some positive constants $c_n$ and $C_n$.
They are defined recursively for $1\leq n\leq N$ by
\be
V_N(s,y,z^i,\gamma_i):=\exp\bigl(\gamma_i F(s,y,z^i)\bigr), \nn
\ee
and
\be
\label{th-t1-eq2}
\begin{split}
V_{n-1}(s,y,z^i,\gamma_i)&:=p_{n-1}(s,y)\exp\Bigl(-\gamma_i\frac{\beta^N}{\beta^n}\phi^{i,*}_{n-1}(s,y,z^i,\gamma_i)u\Bigr)
\ex^{0,i}[V_n(s\wt{u},Y_n,Z_n^i,\gamma_i)|y,z^i,\gamma_i]\\
&+q_{n-1}(s,y)\exp\Bigl(-\gamma_i\frac{\beta^N}{\beta^n}\phi^{i,*}_{n-1}(s,y,z^i,\gamma_i)d\Bigr)
\ex^{0,i}[V_n(s\wt{d},Y_n,Z_n^i,\gamma_i)|y,z^i,\gamma_i].
\end{split}
\ee
\end{theorem}
\begin{remark*}
The function $(1/\gamma_i)\log V_n$ represents the effective liability at $t=t_n$.
\end{remark*}
\begin{proof}
See Appendix~\ref{A-th-t1}.
\end{proof}

\subsection{Mean-field equilibrium under stochastic order flow}
Our goal is to find a set of 
transition probability functions $p_n(s,y), (s,y)\in \cals_n\times \caly_n, 0\leq n\leq N-1$,
that clears the market by matching demand and supply at every node.
To ensure generality, we also incorporate an external stochastic order flow, $L_{n-1}(S_{n-1},Y_{n-1})$, 
which represents the aggregate net stock supply per capita at each time $t_{n-1}$, for $1\leq n\leq N$.
The external order flow serves to model the aggregate contribution from other populations.
In particular, it can be used to represent the aggregate net supply from individual investors, which is often 
difficult to model by rigorous optimizations,  or the supply from a major financial institution
such as a central bank.
\begin{assumption}
\label{assumption-T3}
For every $1\leq n\leq N$,  $L_{n-1}:\cals_{n-1}\times \caly_{n-1}\rightarrow \mbb{R}$
is a bounded measurable function. \footnote{Since the domain $\cals_{n-1}\times \caly_{n-1}$ is finite, the boundedness assumption is redundant. 
However, we make it explicit for clarity. We keep this convention throughout the paper.} 
\end{assumption}
\begin{definition}
We say that the system is in the mean-field equilibrium if 
\be
\lim_{N_p\rightarrow \infty}\frac{1}{N_p}\sum_{i=1}^{N_p}\phi^{i,*}_{n-1}(S_{n-1},Y_{n-1},Z_{n-1}^i,\gamma_i)=L_{n-1}(S_{n-1},Y_{n-1}), \nn
\ee
$\mbb{P}$-a.s.\! for every $1\leq n\leq N$ with $\phi^{i,*}_{n-1}$ defined by $(\ref{th-t1-eq1})$.
\end{definition}

Since the idiosyncratic factors $(Z^i,\gamma_i), i\in \mbb{N}$ are independent and identically distributed (i.i.d.),  
and also independent of the process $(S,Y)$, the above condition for the mean-field equilibrium is equivalently expressed as 
\be
\label{def-mfe-t}
\ex^1\bigl[\phi^{1,*}_{n-1}(s,y,Z_{n-1}^1,\gamma_1)\bigr]=L_{n-1}(s,y)
\ee
for every $(s,y)\in \cals_{n-1}\times \caly_{n-1}$, $1\leq n\leq N$. 
Under the mean-field equilibrium, the excess demand/supply per capita converges to zero as the population size $N_p$ tends to infinity. 
Our first main result is then established as follows.
One can observe a conditional McKean-Vlasov nature, where the transition probabilities
depend on the distribution of idiosyncratic factors conditioned on the realization of common shocks $(S,Y)$,
as expected from the result in \cite{Fujii, Fujii-Sekine1}.

\begin{remark}
We often consider the baseline case $L_{n-1} \equiv 0$, which corresponds to the net zero position among the agents.
The special case $L_{n-1}(s,y)=N^\# s$, $1\leq n\leq N$,  corresponds to the situation 
where the supply is given by a constant number of shares $N^{\#}$ per agent. 
In a closed market setting, this implies that the net initial number of shares per capita is  $N^\#$.
Note that, in our formulation, $L$ represents the monetary value of the supply per capita.
\end{remark}

\begin{theorem}
\label{th-t2}
Let Assumptions~\ref{assumption-T1}, \ref{assumption-T2} and \ref{assumption-T3} be in force.
Then there exists a unique mean-field equilibrium. The associated transition probabilities of the stock price are given  by
\be
\label{th-t2-eq1}
\begin{split}
&p_{n-1}(s,y):=\mbb{P}^0\Bigl(S_n=\wt{u}S_{n-1}|(S_{n-1},Y_{n-1})=(s,y)\Bigr) \\
&=(-d)\Big/ \Bigl\{ u \exp\Bigl(\frac{1}{\ex^1[1/\gamma_1]}\Bigl[\ex^1\Bigl(\frac{\log f_{n-1}(s,y,Z_{n-1}^1,\gamma_1)}{\gamma_1}\Bigr)
-(u-d)\frac{\beta^N}{\beta^n}L_{n-1}(s,y)\Bigr]\Bigr)-d\Bigr\},
\end{split}
\ee
for every $(s,y)\in \cals_{n-1}\times\caly_{n-1}$, $1\leq n\leq N$. 
Here, the functions $f_{n-1}:\cals_{n-1}\times \caly_{n-1}\times \calz_{n-1}\times \Gamma \rightarrow \mbb{R}$, $1\leq n\leq N$,
are given by the backward induction in Theorem~\ref{th-t1}, with the transition probabilities replaced by those given above
at each step.
Under the above transition probabilities, the optimal strategy of agent-$i$ is given by, for each $(s,y,z^i,\gamma_i)\in \cals_{n-1}\times \caly_{n-1}
\times \calz_{n-1}\times\Gamma$,
\be
\label{th-t2-eq3}
\begin{split}
\phi_{n-1}^{i,*}(s,y,z^i,\gamma_i)&=\frac{1}{(u-d)}\frac{\beta^n}{\beta^N}\Bigl\{ \frac{\log f_{n-1}(s,y,z^i,\gamma_i)}{\gamma_i}\\
&\quad -\frac{1/\gamma_i}{\ex^1[1/\gamma_1]}\ex^1\Bigl[\frac{\log f_{n-1}(s,y,Z_{n-1}^1,\gamma_1)}{\gamma_1}\Bigr]\Bigr\}+\frac{1/\gamma_i}{\ex^1[1/\gamma_1]}L_{n-1}(s,y). 
\end{split}
\ee
Moreover, there exists some positive constant $\calc_{n-1}$ such that
\be
\label{th-t2-eq2}
\ex\Bigl|\frac{1}{N_p}\sum_{i=1}^{N_p}\phi^{i,*}_{n-1}(S_{n-1},Y_{n-1}, Z_{n-1}^i,\gamma_i)-L_{n-1}(S_{n-1},Y_{n-1})\Bigr|^2 \leq \frac{\calc_{n-1}}{N_p}
\ee
for every $1\leq n\leq N$, which gives the convergence rate in the large population limit.
\end{theorem}
\begin{proof}
See Appendix~\ref{A-th-t2}.
\end{proof}

\begin{remark}[On the choice of exponential utility]
Before considering the implications of these findings, we offer a brief remark on our choice of exponential utility.
The most important characteristic of the exponential utility is that the optimal trading strategy $\phi^{i,*}$ is independent 
of the wealth process $X^i$ as shown in $(\ref{th-t1-eq1})$.  This property is crucial for solving the mean-field 
condition in Theorem~\ref{th-t2}. Indeed, if other utilities were adopted,  the optimal position $\phi^{i,*}_n$ would generally depend on the agent's
wealth $X_n^i$. The equilibrium condition $(\ref{def-mfe-t})$ would then yield a complex 
fixed-point problem involving the forward $X^i$ and backward $\phi^{i,*}$ processes.
The exponential utility also allows the agents to hold negative net wealth.
This is quite advantageous for modeling financial institutions that naturally hold substantial portfolios of derivatives, 
the values of which can be negative with nonzero probabilities.
Although the constant absolute risk aversion (CARA) is a clear limitation, 
given that most financial institutions routinely revise their risk management objectives, 
the restriction of CARA for a relatively short-term horizon $T$ seems to be a reasonable approximation.
The same observations hold for the recursive utility of exponential-type, which we discuss in Section~\ref{sec-recursive}.
\end{remark}

\subsection{Some implications}
\label{sec-implications}

Let us discuss some implications of the results obtained in this section. 
A key advantage of our closed-form solution is the explicit dependence of the transition probabilities $(p_{n-1}(s,y))$
on the macroeconomic factor $Y$. 
This allows us to clearly analyze how changes in the macroeconomic conditions affect
the equilibrium price distribution and consequently the excess return. 

In order to understand the qualitative behavior of our model, 
assume first that there is no external order flow, i.e.,  $L\equiv 0$ for simplicity.
Recalling that the risk-neutral probability of the up-move at each node is $p^{\mbb{Q}}=(-d)/(u-d)$, 
one can see from $(\ref{th-t2-eq1})$ that $p_{n-1}(s,y)>p^{\mbb{Q}}$ (i.e. a positive excess return at this node)
occurs if and only if \be \ex^1[\log(f_{n-1}(s,y,Z_{n-1}^1,\gamma_1))/\gamma_1]<0.\nn \ee 
This happens if $V_n$ is a decreasing function of the stock price $S_n$,
where $V_n$ represents the effective liability at $t=t_n$ derived from  the terminal liability  $F$.
In Section~\ref{sec-numerical}, we shall confirm this point 
with numerical examples.  
The corresponding situation occurs when the agents' liability (or, the negative of their endowment) decreases as the stock price increases. 
In this case, adding to the long position in the stock increases the risk in the same direction as the liability's exposure, 
and hence the agents require a higher risk premium.
Therefore, for a liability whose size varies countercyclically with the stock price, 
the higher the leverage of the financial and investment firms, the higher the risk premium demanded.
Suppose, on the other hand, that the agents' liability increases when the stock price goes up. 
For example, imagine that agents have a net short position in call options on the stock. 
Then, the agents have a strong incentive to increase their long position in the same stock (to hedge the options), and hence may accept even a negative risk premium.

As one can see from $(\ref{th-t2-eq1})$, it is not necessary to add idiosyncratic shocks to significantly influence the size of excess return,
which is mainly determined by the sensitivity of the liability to the stock price.
However, the absence of idiosyncratic shocks gives rise to a very unrealistic market structure where there is no trade among the agents.
From the expression of the optimal position in $(\ref{th-t2-eq3})$, 
we can observe that the trading volume per capita 
$\ex[|\phi^{1,*}_n|^2]^\frac{1}{2}$ in the market is primarily governed by the variation of the idiosyncratic factors.
Note that, by the definition of the mean-field equilibrium, $\ex[\phi^{1,*}_n]=0, \forall i\in \mbb{N}$ when $L_n=0$.
Therefore, $\ex[|\phi^{1,*}_n|^2]^\frac{1}{2}$ gives the standard deviation of the stock position among the agents
at $t_n$.

In addition to the condition $L\equiv 0$, let us also suppose that the function $F$ is independent of the stock price.
We then have $f_{N-1}=1$ since $V_N$ is $S_N$-independent and thus $\phi^{i,*}_{N-1}=0$ by $(\ref{th-t2-eq3})$.
This in turn makes $V_{N-1}$ independent from $S_{N-1}$. 
In this way, a simple induction shows that $f_{n-1}=1$ for every $1\leq n\leq N$
and the equilibrium price distribution becomes equal to the one in the risk-neutral measure.
In this case,  there is no trade in the market although each agent has different risk aversion, 
which corresponds to the classical (but uninteresting) example of the representative agent with  CARA utility.

Finally, let us turn on the external order flow. It is clearly seen from $(\ref{th-t2-eq1})$ that the positive inflow  $L>0$
to the stock market 
increases the equilibrium risk premium. This may sound slightly counter-intuitive since we think a big sell-off in the stock should lead to 
a sharp decline in the stock price. In order to understand that there is no contradiction, 
it is important to recall that what we have found above is the transition probabilities so that there exists equilibrium.
As a common characteristic of price-formation frameworks, our model does not argue the performance of a stock price
given the business performance of its issuer. Rather,  it provides the excess return (and more generally price distribution)
required by the agents to support the existence of the market equilibrium.
If there is positive supply of the stock, the agents must accept larger long position (and hence larger risk) in the stock to maintain 
the balance of demand and supply; consequently,  the agents require a higher risk premium to compensate  for this additional risk.  
If the risk premium is not high enough, there would be no equilibrium and thus the stock market might  crash.
If the required risk premium turns out to be unrealistically high,  then one can infer that 
the stock market cannot support such a large external supply.
The same analysis can be done for the sustainability of the stock equilibrium under highly levered financial firms.

\section{Recursive utility with path-dependent financial items}
\label{sec-recursive}

In the previous section, we obtained mean-field equilibrium by choosing an appropriate set of transition probabilities in the form of $p_n(s,y)$. 
Suppose now that the stochastic liability (or the negative of the endowment) $F$ is dependent 
not only on the terminal stock price $S_N$ but also on the stock-price history $(S_n)_{n=0}^N$, 
which is just as plausible.  In this case,  a quick inspection of the proofs for Theorems~\ref{th-t1} and \ref{th-t2} shows that the transition probabilities of the simple form $p_n(s,y)$ cannot clear the market anymore.  
This strongly suggests that we need path-dependence also in  the transition probabilities. 
We also want to examine whether we can include cash spending (i.e.~nominal consumption) and to analyze its impact on the excess return.
In this section, we shall thus adopt recursive utility that incorporates
standard time-separable utility over nominal consumptions as its special case. We  include a path-dependent terminal liability as well as
path-dependent incremental endowments in the model for generality.

\subsection{The setup and notation}
\label{sec-setup-recursive}
In this section, for each $i\in \mbb{N}$, 
we enlarge the probability space $(\Omega^i, \calf^i, (\calf^i_{t_n})_{n=0}^N, \mbb{P}^i)$  so that
it supports $(\xi_i, \gamma_i, \zeta_i, \psi_i, \del_i)$ as $\calf^i_0$-measurable random variables, 
in addition to $(\calf_{t_n}^i)_{n=0}^N$-adapted stochastic process $(Z_n^i=Z^i(t_n))_{n=0}^N$.
Here, $\zeta_i$ is the coefficient of absolute risk aversion for cash spending and the parameter $\psi_i$ is used 
to control the importance of the continuation utility relative to the current spending.
$\del_i$ denotes the coefficient of time preference.  We introduce an $\calf_0^i$-measurable 4-tuple
$\vr_i:=(\gamma_i,\zeta_i, \psi_i, \del_i)$ for simpler notation. 
Let us also introduce the symbol $\bS^n:=(S_0, S_1,\ldots, S_n)$ to denote a stock-price trajectory
and $\bs^n=(s_0,\ldots, s_n)\in \cals^n$ as its specific realization. 
For $\bs\in \cals^{n-1}$, we also use the symbols $(\bs \wt{u})^n:=(\bs^{n-1}, s_{n-1}\wt{u})\in \cals^n$
and $(\bs \wt{d})^n:=(\bs^{n-1},s_{n-1}\wt{d})\in \cals^n$. 
As in the last section,  we shall use the expressions such as 
$\ex^{0,i}[ \cdot |\bs, y,z^i,\vr_i]$ for $(\bs,y,z^i)\in \cals^{n-1}\times \caly_{n-1}\times \calz_{n-1}$
to denote the conditional expectation $\ex^{0,i}[ \cdot |\bS^{n-1}=\bs, Y_{n-1}=y,Z_{n-1}^i=z_i, \vr_i=\vr_i]$,
where, with a slight abuse of notation,  the same symbol is used for a realization of the $\calf^i_0$-measurable random variable $\vr_i$.
Except these points, we will use the same setup and notation  given in Section~\ref{sec-setup}. 
In particular, we impose the finite state space condition only on $(S_n)$ and $(Y_n)$.
Now, let us update Assumption~\ref{assumption-T1} for this section.

\begin{assumption}
\label{assumption-R1}
{\rm (i):} $\wt{u}$ and $\wt{d}$ are real constants satisfying
\be
0<\wt{d}<\exp(r\Del)<\wt{u}<\infty. \nn
\ee 
{\rm (ii):} The variables $(\xi_i,\gamma_i, \zeta_i, \psi_i, \del_i, (Z_n^i)_{n=0}^N)$ are identically distributed across all agents $i=1,2,\ldots$.\\
{\rm (iii):} There exist real constants $\ul{\xi}, \ol{\xi}$,  $\ul{\gamma}, \ol{\gamma}$, $\ul{\zeta},\ol{\zeta}$, $\ul{\psi},\ol{\psi}$
and $\ul{\delta},\ol{\delta}$ so that for every $i\in \mbb{N}$,
\be
\begin{split}
&\xi_i\in [\ul{\xi},\ol{\xi}]\subset \mbb{R}, \\
&\vr_i:=(\gamma_i,\zeta_i, \psi_i,\delta_i) \in \Gamma:=[\ul{\gamma},\ol{\gamma}]\times [\ul{\zeta},\ol{\zeta}]
\times [\ul{\psi},\ol{\psi}]\times [\ul{\delta},\ol{\delta}]
\subset (0,\infty)^4. \nn
\end{split}
\ee 
{\rm (iv):} For each $i$, $(Z_n^i)_{n=0}^N$ is a Markov process i.e.~$\mbb{E}^i[f(Z_n^i)|\calf_{t_k}^i]=\ex^i[f(Z_n^i)|Z_k^i]$
for every bounded measurable function $f$  on $\calz_n$ and $k\leq n$.\\
{\rm (v):} $(Y_n)_{n=0}^N$ is a Markov process i.e.~$\mbb{E}^0[f(Y_n)|\calf_{t_k}^0]=\ex^0[f(Y_n)|Y_k]$
for every bounded measurable function $f$ on $\caly_n$ and $k\leq n$. \\
{\rm (vi):} The transition probabilities of $(S_n)_{n=0}^N$ satisfy, for every $0\leq n\leq N-1$, a.s.,
\be
\begin{split}
\mbb{P}^0(S_{n+1}=\wt{u}S_{n}|\calf_{t_n}^0)&=\mbb{P}^0(S_{n+1}=\wt{u}S_n|\bS^n,Y_n)=:p_{n}(\bS^{n},Y_{n}), \\
\mbb{P}^0(S_{n+1}=\wt{d}S_{n}|\calf_{t_n}^0)&=\mbb{P}^0(S_{n+1}=\wt{d}S_n|\bS^n,Y_n)=:q_{n}(\bS^{n},Y_{n}), \nn
\end{split}
\ee
where $p_{n}, q_n~(:=1-p_n):\cals^{n}\times \caly_n\rightarrow \mbb{R}, ~0\leq n\leq N-1$ are bounded measurable functions
satisfying 
\be
0<p_n(\bs,y), q_n(\bs,y)<1 \nn
\ee
for every $(\bs,y)\in \cals^n\times \caly_n$.
\end{assumption}
\noindent
Under the above assumptions, we have,  instead of $(\ref{Y-conditional-1})$, the relation
\be
\label{Y-conditional-2}
\ex^0\bigl[f(S_{n+1})g(Y_{n+1})|\calf_{t_n}^0]=\ex^0[f(S_{n+1})|\bS^n,Y_n]\ex^0[g(Y_{n+1})|Y_n]~{\rm a.s.,} \quad 0\leq n\leq N-1,
\ee
for any bounded measurable functions $f:\cals_{n+1}\rightarrow \mbb{R}$ and 
$g:\caly_{n+1}\rightarrow \mbb{R}$.

\begin{remark}
As in the last section, the condition $0<p_n(\bs, y), q_n(\bs,y)<1, \forall (\bs,y)\in \cals^n\times \caly_n, 0\leq n\leq N-1$
guarantees the equivalence of $\mbb{P}^0\circ S^{-1}$ and $\mbb{Q}\circ S^{-1}$. Hence the
system is arbitrage free.
\end{remark}

\subsection{The individual optimization problem}
\label{sec-rec-individual}
In this section, as previously mentioned, we assume that each agent-$i$ engages in self-financing trading 
using the money-market account and the risky stock, while also spending some cash at the beginning of each period. Moreover,
they receive a stochastic endowment at each time $t_n, 1\leq n\leq N$.
Thus the  wealth of the agent-$i$,  $(X^i_n:=X^i(t_n))_{n=0}^N$, follows the dynamics:
\be
\label{eq-rec-wealth}
\begin{split}
X_{n+1}^i&=\exp(r\Del)(X_n^i-c_n^i \Del-\phi_n^i)+\phi_n^i \wt{R}_{n+1}+g_{n+1}(\bS^{n+1},Y_{n+1},Z_{n+1}^i)\\
&=\beta(X_n^i-c_n^i \Del)+\phi_n^i R_{n+1}+g_{n+1}(\bS^{n+1},Y_{n+1},Z_{n+1}^i), 
\end{split}
\ee 
where $X_0^i=\xi_i$. Recall that $R_n:=\wt{R}_n-\exp(r\Del)$. Here, $c_n^i , 0\leq n\leq N-1$ denotes the cash spending at $t_n$.
For the scaling purposes in the discrete-time model, this value is defined such that the actual 
cash used is $c_n^i \Del$. 
$g_n(\bS^n,Y_n,Z_n^i), 1\leq n\leq N$ is the stochastic endowment (i.e., income originating from the agent's other business lines) 
paid at $t_n$, which  is dependent on the stock-price trajectory $\bS^n$
in addition to the common and the idiosyncratic shocks $(Y_n, Z_n^i)$. 
 Since we do not restrict $g_n$ to be positive,
it can also represent incremental liability incurred at $t_n$ when negative.

\bigskip
We suppose that the $(\calf^{0,i}_{t_n})_{n=0}^N$-adapted process of utilities $(U_n^i)_{n=0}^N$ is  defined recursively by
\be
\label{def-RU}
\begin{split}
&U_n^i:=-\frac{1}{\zeta_i}\log\Bigl\{\exp\bigl(-\zeta_i c_n^i\bigr)\Del+\del_i \exp\Bigl(\frac{\psi_i}{\gamma_i}\log\bigl(\ex^{0,i}[e^{-\gamma_i U_{n+1}^i}|\calf_{t_n}^{0,i}]
\bigr)\Bigr)\Bigr\}, \\
&\leftrightarrow 
\exp(-\zeta_i U_n^i)=\exp(-\zeta_i c_n^i)\Del+\del_i \exp\Bigl(\frac{\psi_i}{\gamma_i}\log\bigl(\ex^{0,i}[e^{-\gamma_i U_{n+1}^i}|\calf_{t_n}^{0,i}]\bigr)\Bigr),
\end{split}
\ee
with the terminal condition
\be
U_N^i:=X_N^i-F(\bS^N,Y_N,Z_N^i). \nn
\ee
Here, $F(\bS^N, Y_N,Z_N^i)$ denotes the terminal liability as in the last section, but now dependent on the price trajectory $\bS^N$.
Each agent-$i$ is supposed to solve the optimization problem
\be
\label{problem-R1}
\sup_{(\phi_n^i,c_n^i)_{n=0}^{N-1}\in \mbb{A}^i} U^i_0, 
\ee
over the admissible space defined by
\be
\mbb{A}^i:=\bigl\{(\phi_n^i, c_n^i)_{n=0}^{N-1}: (\phi_n^i, c_n^i) \text{ is an $\calf_n^{0,i}$-measurable $\mbb{R}^2$-valued random variable} \bigr\}. \nn
\ee
For simplicity, we do not restrict $(c_n^i)$ to non-negative values. One may interpret negative spending  
as positive income from costly labor for the corresponding period.

\begin{assumption}
\label{assumption-R2}
{\rm (i):} The function $F:\cals^N\times \caly_N\times \calz_N\rightarrow \mbb{R}$ is measurable and bounded.\\
{\rm (ii):} For every $1\leq n\leq N$, the function $g_n:\cals^n\times\caly_n\times \calz_n\rightarrow \mbb{R}$
is measurable and bounded. \\
{\rm (iii):}  Every agent is a price-taker in the sense that they consider the stock price process (and hence its transition 
probabilities specified in Assumption~\ref{assumption-R1} (vi)) to be exogenously given  by the collective actions 
of the others and unaffected by the agent's own trading strategies.
\end{assumption}

Before going into the details, let us consider the special case: $\zeta_i=\gamma_i=\psi_i$. Then we have
\be
\begin{split}
\exp({-\zeta_i U_n^i})&=\exp({-\zeta_i c_n^i})\Del +\del_i \ex^{0,i}[\exp({-\zeta_i U_{n+1}^i})|\calf_{t_n}^{0,i}]\\
&=\exp({-\zeta_i c_n^i})\Del +\ex^{0,i}\bigl[\del_i \exp({-\zeta_ic_{n+1}^i})\Del |\calf_{t_n}^{0,i}\bigr]
+\del_i^2\ex^{0,i}[\exp({-\zeta_i U_{n+2}^i})|\calf_{t_n}^{0,i}] \\
&=\cdots =\exp({-\zeta_i c_n^i})\Del+\ex^{0,i}\Bigl[\sum_{k=n+1}^{N-1} \del_i^{k-n} \exp({-\zeta_i c_k^i})\Del+\del_i^{N-n}
\exp({-\zeta_i U_N^i})|\calf_{t_n}^{0,i}\Bigr], \nn
\end{split}
\ee
which thus corresponds to the standard time-separable utility over nominal consumptions with the terminal liability $F$. 
One can see that the parameter $\psi_i$
determines the relative importance of the continuation utility.
We now derive the optimal strategy for each agent with respect to  the above-defined recursive utility.

\begin{theorem}
\label{th-R1}
Let Assumptions~\ref{assumption-R1} and \ref{assumption-R2} be in force.
Then the problem $(\ref{problem-R1})$ has an a.s.\! unique optimal solution $(\phi_{n-1}^{i,*}, c_{n-1}^{i,*})_{n=1}^{N}$,
where $(\phi_{n-1}^{i,*})_{n=1}^N$ and $(c_{n-1}^{i,*})_{n=1}^N$ are bounded processes defined by
measurable functions $\phi_{n-1}^{i,*}:\cals^{n-1}\times \caly_{n-1}\times \calz_{n-1}\times \Gamma\rightarrow \mbb{R}$
and $c_{n-1}^{i,*}:\mbb{R}\times \cals^{n-1}\times \caly_{n-1}\times \calz_{n-1}\times \Gamma\rightarrow \mbb{R}$
such that $\phi_{n-1}^{i,*}:=\phi^{i,*}_{n-1}(\bS^{n-1},Y_{n-1},Z_{n-1}^i,\vr_i)$
and $c_{n-1}^{i,*}:=c^{i,*}_{n-1}(X_{n-1}^i, \bS^{n-1}, Y_{n-1}, Z_{n-1}^i, \vr_i)$ respectively,
where, for each $(x^i, \bs, y, z^i,\vr_i)\in \mbb{R}\times \cals^{n-1}\times \caly_{n-1}\times \calz_{n-1}\times \Gamma$,
\be
\begin{split}
\label{th-R1-phinm1}
\phi_{n-1}^{i,*}(\bs,y, z^i,\vr_i):=\frac{1}{\gamma_i \eta_n^i (u-d)}\Bigl\{ \log\Bigl(-\frac{p_{n-1}(\bs,y)u}{q_{n-1}(\bs,y)d}\Bigr)
+\log\bigl(f_{n-1}(\bs, y, z^i,\vr_i)\bigr)\Bigr\}, 
\end{split}
\ee
\be
\begin{split}
\label{th-R1-cnm1}
c_{n-1}^{i,*}(x^i,\bs,y,z^i,\vr_i):=\frac{\psi_i \eta_n^i \beta}{\zeta_i+\Del \psi_i \eta_n^i \beta}x^i
-\frac{1}{\zeta_i+\Del \psi_i \eta_n^i\beta}\log\Bigl\{ \frac{\del_i \psi_i \eta_n^i \beta}{\zeta_i}\exp\Bigl(
\frac{\psi_i}{\gamma_i}\log\wt{V}_{n-1}(\bs,y,z^i,\vr_i)\Bigr)\Bigr\}.
\end{split}
\ee
For every $1\leq n\leq N$, $f_{n-1}, \wt{V}_{n-1}:\cals^{n-1}\times \caly_{n-1}\times \calz_{n-1}\times \Gamma\rightarrow \mbb{R}$ 
and $V_n:\cals^n\times\caly_n\times \calz_n\times \Gamma\rightarrow \mbb{R}$ are measurable functions.
$f_{n-1}$ and $\wt{V}_{n-1}$ satisfy the uniform bounds $0<c_n \leq f_{n-1}, \wt{V}_{n-1} \leq C_n<\infty$,
while $V_n$ is bounded $(|V_n|\leq C_n)$, for some positive constants $c_n$ and  $C_n$. They are defined recursively for $1\leq n\leq N$
by
\be
\label{th-R1-fnm1}
f_{n-1}(\bs,y,z^i,\vr_i):=\frac{\ex^{0,i}\bigl[\exp\bigl(\gamma_i [V_n((\bs \wt{u})^n, Y_n, Z_n^i,\vr_i)-\eta_n^i g_n((\bs \wt{u})^n, Y_n,Z_n^i)]\bigr)|y,z^i,\vr_i\bigr]}
{\ex^{0,i}\bigl[\exp\bigl(\gamma_i [V_n((\bs \wt{d})^n, Y_n,Z_n^i,\vr_i)-\eta_n^i g_n((\bs \wt{d})^n, Y_n,Z_n^i)]\bigr)|y,z^i,\vr_i\bigr]},
\ee
and 
\be
\label{th-R1-Vnm1}
\begin{split}
&V_{n-1}(\bs, y,z^i,\vr_i):=\frac{\eta_{n-1}^i}{\eta_n^i\gamma_i\beta}\log \wt{V}_{n-1}(\bs,y,z^i,\vr_i)
+\frac{1}{\zeta_i+\Del \psi_i \eta_n^i\beta}\log\Bigl(\frac{\del_i\psi_i\eta_n^i\beta}{\zeta_i}\Bigr)-\frac{1}{\zeta_i}\log(\eta_{n-1}^i), 
\end{split}
\ee
where 
\be
\begin{split}
\label{th-R1-Vtildenm1}
&\wt{V}_{n-1}(\bs,y,z^i,\vr_i)\\
&~:=p_{n-1}(\bs,y)e^{-\gamma_i \eta_n^i \phi_{n-1}^{i,*} u}
\ex^{0,i}\bigl[\exp(\gamma_i[V_n((\bs \wt{u})^n,Y_n,Z_n^i,\vr_i)-\eta_n^i g_n((\bs \wt{u})^n,Y_n,Z_n^i)])|y,z^i,\vr_i\bigr] \\
&~+q_{n-1}(\bs,y)e^{-\gamma_i \eta_n^i \phi_{n-1}^{i,*}d}
\ex^{0,i}\bigl[\exp(\gamma_i[V_n((\bs \wt{d})^n,Y_n,Z_n^i,\vr_i)-\eta_n^i g_n((\bs \wt{d})^n,Y_n,Z_n^i)])|y,z^i,\vr_i\bigr],  
\end{split}
\ee
starting from the terminal condition $V_N(\bS^N, Y_N,Z_N^i,\vr_i):=F(\bS^N,Y_N,Z_N^i)$. $(\eta_n^i)_{n=0}^N$ are strictly positive and bounded
$\calf_0^i$-measurable random variables given by the recursive relation:
\be
\label{th-R1-etanm1}
\eta_{n-1}^i:=\frac{\psi_i \eta_n^i \beta}{\zeta_i+\Del \psi_i \eta_n^i\beta}, \quad \eta_N^i\equiv 1. 
\ee
\end{theorem}
\begin{remark*}
Similar to Theorem~\ref{th-t1}, $V_n$ represents  the effective liability at $t_n$.
\end{remark*}
\begin{proof}
See Appendix~\ref{A-th-R1}.
\end{proof}

\begin{remark}[On time inconsistency]
\label{rem-time-inconsistency}
It is well known that the optimization problem for utilities becomes, in general,  time-inconsistent
if the associated coefficients are time-dependent, such as $( \gamma_i(t), \zeta_i(t), \del_i(t), \psi_i(t))$.
However, the solution $(\phi^{i,*}_n, c^{i,*}_n)_{n=0}^{N-1}$,
when derived via backward induction as we have done in this section,
produces the sub-game perfect Nash equilibrium (SPNE). 
This SPNE solution is consistent with the interpretation that an independent 
agent is responsible for the optimization in each interval $[t_n,t_{n+1}]$, 
and they do not commit to past decisions made by previous agents but act optimally believing that 
future agents will do  the same.
The  backward induction solution $(\phi^{i,*}_n,c^{i,*}_n)_{n=0}^{N-1}$
yields a Nash equilibrium for these agents, which is an important characteristic of 
any games defined on trees over finite intervals.
Considering that most financial firms revise their risk and budgetary targets periodically (possibly under new managers), 
the concept of SPNE aligns well with the real-world practices. Therefore, even if we deal with time-inconsistent problem
with time-dependent coefficients $(\gamma_i(t), \zeta_i(t), \del_i(t), \psi_i(t))$, our scheme still provides
a meaningful result. This is an additional advantage of adopting the binomial tree approach, 
which sidesteps the complexities often encountered in continuous-time settings.
Unfortunately, however, if the coefficients follow general stochastic processes not measurable by $\calf_0^i$, 
explicit solvability would be lost.
\end{remark}

\subsection{Mean-field equilibrium among the agents with recursive utilities}
\label{sec-mfg-recursive}
Finally, as a main goal of this section, we shall derive a set of transition probabilities of the stock price so that the mean-field equilibrium holds
among the agents with recursive utilities. As before, we incorporate the existence of stochastic external order flow $L_{n}$ at each $t_n$, 
but  it is now  allowed to be path-dependent on the stock price:
\begin{assumption}
\label{assumption-R3}
For every $1\leq n\leq N$, $L_{n-1}:\cals^{n-1}\times \caly_{n-1}\rightarrow \mbb{R}$ is a bounded measurable function.
\end{assumption} 

\begin{definition}
\label{def-mfe-rec}
We say that the system is in the mean-field equilibrium if
\be
\lim_{N_p\rightarrow \infty}\frac{1}{N_p}\sum_{i=1}^{N_p}\phi^{i,*}_{n-1}(\bS^{n-1},Y_{n-1},Z_{n-1}^i,\vr_i)=L_{n-1}(\bS^{n-1},Y_{n-1}), \nn
\ee
$\mbb{P}$-a.s. for every $1\leq n\leq N$ with $\phi^{i,*}_{n-1}$ defined by $(\ref{th-R1-phinm1})$.
\end{definition}
Since $(Z^i,\vr_i), i\in \mbb{N}$ are independent, identically distributed,  and also independent of the process $(S,Y)$, the above condition for the mean-field equilibrium
can be represented by
\be
\label{rec-clearing}
\ex^1\bigl[\phi^{1,*}_{n-1}(\bs,y,Z^1_{n-1},\vr_1)\bigr]=L_{n-1}(\bs, y)
\ee
for every $(\bs, y)\in \cals^{n-1}\times\caly_{n-1}$, $1\leq n\leq N$.
It is now straightforward to derive the counterpart of Theorem~\ref{th-t2}.

\begin{theorem}
\label{th-R2}
Let Assumptions~\ref{assumption-R1}, \ref{assumption-R2} and \ref{assumption-R3} be in force.
Then there exists a unique mean-field equilibrium.  The associated transition probabilities of the stock price are given by
\be
\begin{split}
&p_{n-1}(\bs,y):=\mbb{P}^0\Bigl(S_n=\wt{u}S_{n-1}|(\bS^{n-1},Y_{n-1})=(\bs,y)\Bigr)\\
&=(-d)\Big/ \Bigl\{ u\exp\Bigl(\frac{1}{\ex^1[1/(\gamma_1\eta_n^1)]}\Bigl[
\ex^1\Bigl(\frac{\log (f_{n-1}(\bs,y,Z^1_{n-1},\vr_1))}{\gamma_1\eta_n^1}\Bigr)-(u-d)L_{n-1}(\bs,y)\Bigr]\Bigr)-d\Bigr\} \nn
\end{split}
\ee
for every $(\bs,y)\in \cals^{n-1}\times \caly_{n-1}$, $1\leq n\leq N$. 
Here, the functions $f_{n-1}:\cals^{n-1}\times \caly_{n-1}\times \calz_{n-1}\times \Gamma\rightarrow \mbb{R}$,
$1\leq n\leq N$, are given by the backward induction in Theorem~\ref{th-R1}, with the transition 
probabilities replaced by those given above at each step.
Under the above transition probabilities, the optimal strategy of agent-$i$ is given by
\be
\label{th-R2-phinm1}
\begin{split}
\phi^{i,*}_{n-1}(\bs,y,z^i,\vr_i)&=\frac{1}{(u-d)}\Bigl\{\frac{\log f_{n-1}(\bs,y,z^i,\vr_i)}{\gamma_i \eta_n^i}\\
&\quad-\frac{1/(\gamma_i \eta_n^i)}{\ex^1[1/(\gamma_1\eta_n^1)]}\ex^1\Bigl(\frac{\log f_{n-1}(\bs,y,Z_{n-1}^1,\vr_1)}{\gamma_1\eta_n^1}\Bigr)
\Bigr\}+\frac{1/(\gamma_i \eta_n^i)}{\ex^1[1/(\gamma_1\eta_n^1)]}L_{n-1}(\bs,y).
\end{split}
\ee
Moreover, there exists some positive constant $\calc_{n-1}$ such that
\be
\ex\Bigl|\frac{1}{N_p}\sum_{i=1}^{N_p}\phi^{i,*}_{n-1}(\bS^{n-1},Y_{n-1},Z_{n-1}^i,\vr_i)-L_{n-1}(\bS^{n-1},Y_{n-1})\Bigr|^2\leq
\frac{\calc_{n-1}}{N_p} \nn
\ee
for every $1\leq n\leq N$, which gives the convergence rate in the large population limit.
\end{theorem}
\begin{proof}
See Appendix~\ref{A-th-R2}
\end{proof}

\begin{remark}
\label{remark-path}
The path-dependent extension for the model in Section~\ref{sec-terminal},
which replaces $F(S_N, Y_N,Z_N^i)$ and  $(L_{n-1}(S_{n-1},Y_{n-1}))_{n=1}^N$
with the path-dependent forms $F(\bS^N, Y_N, Z_N^i)$ and  $(L_{n-1}(\bS^{n-1}, Y_{n-1}))_{n=1}^N$, can be done in exactly the same way.
Specifically, the corresponding results are obtained by substituting the path-dependent states 
$(\bs, (\bs \wt{u})^n, (\bs\wt{d})^n)\in \cals^{n-1}\times \cals^n\times\cals^n$ 
for one-time states $(s,s\wt{u},s\wt{d})\in \cals_{n-1}\times \cals_n\times \cals_n$ in the statements of Theorems~\ref{th-t1} and \ref{th-t2}.

More generally, it is not difficult to confirm that the path dependence horizon required for the transition probabilities 
matches that for the liability and the incremental endowments.
In particular, if $F$ and $g_n$ in the current recursive utility model depend only on $S_N$ and $S_n$ respectively,  as in Section~\ref{sec-terminal}, 
then the mean-field equilibrium is achieved using transition probabilities of the form $(p_{n-1}(s, y), q_{n-1}(s,y))$ with 
$s\in \cals_{n-1}$; by simply replacing $\bs\in \cals^{n-1}$ with $s\in \cals_{n-1}$, we obtain the corresponding results for 
Theorems~\ref{th-R1} and \ref{th-R2}.
\end{remark}

\subsection{Some implications}
\label{sec-rec-implication}
Let us discuss some implications of the results in this section for the recursive utility.
The key advantage of the explicit dependence of the transition probabilities $(p_{n-1}(s,y))$
on the macroeconomic factor $Y$ still holds in the current case. 
Regarding the relation between the mean-field price distribution and the one in the risk-neutral measure, 
most of the discussions given in Section~\ref{sec-implications} still hold.
Indeed, the situation $p_{n-1}(\bs,y)>p^{\mbb{Q}}$ (i.e., a positive excess return at the node $(\bs, y)$)
occurs if and only if 
\be
\ex^1\bigl[\log f_{n-1}(\bs,y,Z^1_{n-1},\vr_1) /(\gamma_1\eta_n^1)\bigr]<0, \nn
\ee
when there is no external order flow $L_{n-1}\equiv 0$.

As one can see from $(\ref{th-R1-fnm1})$, $f_{n-1}$ now receives contributions
from the incremental endowment $g_n$ in addition to the liability $V_n$.
For example, although the investment funds typically do not have substantial liabilities,  their endowments (fees from their customers) are naturally expected to grow as the stock price goes higher, since these fees are generally proportional to the Assets Under Management (AUM),
which pushes the required excess return to higher values.
If the liability size decreases and the endowment size simultaneously increases as the stock price grows, both effects  will amplify the deviations from the risk-neutral distribution, leading to  a higher excess return.

Since the equilibrium price distribution of  the stock is determined by the need for risk hedging, the relative importance of the continuation utility
with respect to the current nominal consumption is a crucial factor in controlling the size of the risk premium.
From the expressions in $(\ref{th-R1-Vnm1})$ and $(\ref{th-R1-Vtildenm1})$, one can expect that the ratio
\be
\eta_{n-1}^i/\eta_n^i \nn
\ee
is the key value. Using $\beta\simeq 1$ and $\Del \ll 1$, we have
$
\eta_{n-1}^i/\eta_n^i\simeq \psi_i/\zeta_i \nn
$
from $(\ref{th-R1-etanm1})$. Therefore, if 
$\psi_i<\zeta_i$ holds for the majority of agents, we expect that the relative importance of the continuation utility 
quickly decays and we would see  only a small impact from it in the  earlier periods. In this case, significant deviations from the risk-neutral price distribution can be observed only in the later periods,  near the maturity. On the other hand, in the case of $\psi_i\geq\zeta_i$, 
we can expect to see significant deviations throughout the interval. We shall confirm this behavior by numerical examples in Section~\ref{sec-numerical}.

As for the expected trading volume $\ex[|\phi^{1,*}_n|^2]^\frac{1}{2}$, which gives the standard deviation of the stock 
position among the agents at $t=t_n$, the result is consistent with that in Section~\ref{sec-implications}.
The expression for $\phi^{i,*}_{n-1}$ in $(\ref{th-R2-phinm1})$ shows that its size is governed by the 
variation of idiosyncratic factors defined on the space $(\Omega^i, \calf^i,  \mbb{P}^i)$.
It is also quite consistent with our intuition 
that the agents' heterogeneity in idiosyncratic factors is the origin of the trading activity in the market.
Moreover, we can make use of the degrees of freedom in the process $(Z_n^i)$, in particular its volatility,
to obtain the desired trading volume. We shall demonstrate this property by numerical examples in Section~\ref{sec-num-recursive}.

Finally, we comment on the fact that the constant shifts in $F$ and $g_n$, i.e. $F\mapsto F+c$ and $g_n\mapsto g_n+c^\prime$ 
with some constants $c,c^\prime\in \mbb{R}$ do not affect the equilibrium price distribution.
This property can be checked by a simple induction as follows:
By $(\ref{th-R1-fnm1})$, the value of $f_{N-1}$ remains unchanged and so are $p_{N-1}(\bs,y)$ and $\phi^{i,*}_{N-1}$;
the value of $\wt{V}_{N-1}$ is changed only by an $\calf^i_0$-measurable multiplicative factor;
the value of $V_{N-1}$ is simply shifted by an $\calf^i_0$-measurable term;
thus $f_{N-2}$ remains once again unchanged, and so are $p_{N-2}(\bs,y)$ and $\phi^{i,*}_{N-2}$, and so on.
Therefore, the signs of $F$ and $g_n$ can be altered without affecting the equilibrium price distribution. 
Note however that the cash spending is affected by these shifts.

\section{Mean-field equilibrium of multiple populations}
\label{sec-multi-p}
A primary drawback of the previous frameworks lies in their restriction to a single homogeneous population,
where all agents share identical functional forms for terminal liabilities $F$ and incremental endowments $(g_n)_{n=1}^N$.
That is, their heterogeneity is limited to the realizations of the idiosyncratic factor process $Z^i$.
Beyond this structural constraint, risk aversion coefficients and time preferences are expected to have substantially different distributions
across diverse financial entities, such as investment banks, commercial banks, insurance firms,  pension funds,  and other investment funds. 
To address  this fundamental market heterogeneity, we propose a multi-population extension of the mean-field equilibrium studied in the preceding sections.

Let us consider the case where there are $m\in \mbb{N}$ populations. 
For simplicity, we assume that all the agents have recursive utilities of exponential-type as in Section~\ref{sec-recursive}.
We adopt the same model setup and notation as given in Sections~\ref{sec-setup} and \ref{sec-setup-recursive},
in particular those for  $(S_n)_{n=0}^N$ and $(Y_n)_{n=0}^N$. 
However to incorporate the heterogeneity across the $m$ populations, we need to 
introduce population-specific probability spaces for the idiosyncratic factors.
We now introduce a countably infinite number of  complete 
filtered probability spaces $(\Omega^{i,p}, \calf^{i,p}, (\calf^{i,p}_{t_n})_{n=0}^N, \mbb{P}^{i,p})$, $i\in \mbb{N}$
for each population $p$, $p=1,\ldots, m$.
We introduce $(\xi_i^p, \gamma_i^p, \zeta_i^p, \psi_i^p, \del_i^p)$ as $\calf_0^{i,p}$-measurable $\mbb{R}$-valued random variables
and $(Z_n^{i,p}=Z^{i,p}(t_n))_{n=0}^N$ as $(\calf^{i,p}_{t_n})_{n=0}^N$-adapted $d_{Z^p}$-dimensional stochastic process.
As before, we denote the range of the random variable $Z_n^{i,p}$ by $\calz_n^p$.
Importantly, while these variables are assumed to be i.i.d.~copies within each population $p$, their distributions
can vary across the populations. 

With the same probability space $(\Omega^0, \calf^0, (\calf^0_{t_n})_{n=0}^N, \mbb{P}^0)$ as in Section~\ref{sec-setup},
we define
\be
(\Omega,\calf, (\calf_{t_n})_{n=0}^N, \mbb{P}):=(\Omega^0, \calf^0, (\calf^0_{t_n})_{n=0}^N, \mbb{P}^0)
\otimes_{p=1}^m \otimes_{i=1}^\infty (\Omega^{i,p},\calf^{i,p}, (\calf^{i,p}_{t_n})_{n=0}^N, \mbb{P}^{i,p}) \nn
\ee
as the full probability space containing the entire environment of the $m$-population model.
For agent-$i$ in the population $p$ (it will be denoted by agent-$(i,p)$ hereafter), the relevant probability space is given by
\be
(\Omega^{0, (i,p)},\calf^{0, (i,p)}, (\calf_{t_n}^{0,(i,p)})_{n=0}^N, \mbb{P}^{0,(i,p)}):=(\Omega^0, \calf^0, (\calf^0_{t_n})_{n=0}^N, \mbb{P}^0)
\otimes (\Omega^{i,p},\calf^{i,p}, (\calf^{i,p}_{t_n})_{n=0}^N, \mbb{P}^{i,p}). \nn
\ee
Expectations with respect to $\mbb{P}^{i,p}$ and $\mbb{P}^{0,(i,p)}$ are denoted by $\ex^{i,p}[\cdot]$ and $\ex^{0,(i,p)}[\cdot]$, respectively.
We assume the following for the multi-population model:
\begin{assumption}
\label{assumption-Rm1}
{\rm (i):} $\wt{u}$ and $\wt{d}$ are real constants satisfying
$0<\wt{d}<\exp(r\Del)<\wt{u}<\infty$. \\
{\rm (ii):} The variables $(\xi_i^p, \gamma_i^p, \zeta_i^p, \psi_i^p, \del_i^p, (Z_n^{i,p})_{n=0}^N)$ are identically distributed 
across all agents $i=1,2,\ldots$ within each population $1\leq p\leq m$.\\
{\rm (iii):} For each $1\leq p\leq m$, there exist real constants $\ul{\xi}^p, \ol{\xi}^p$,  $\ul{\gamma}^p, \ol{\gamma}^p$, $\ul{\zeta}^p,\ol{\zeta}^p$, $\ul{\psi}^p,\ol{\psi}^p$
and $\ul{\delta}^p,\ol{\delta}^p$ so that for every $i\in \mbb{N}$, 
\be
\begin{split}
&\xi_i^p\in [\ul{\xi}^p,\ol{\xi}^p]\subset \mbb{R}, \\
&\vr_i^p:=(\gamma_i^p,\zeta_i^p, \psi_i^p,\delta_i^p) \in \Gamma^p:=[\ul{\gamma}^p,\ol{\gamma}^p]\times [\ul{\zeta}^p,\ol{\zeta}^p]
\times [\ul{\psi}^p,\ol{\psi}^p]\times [\ul{\delta}^p,\ol{\delta}^p]
\subset (0,\infty)^4. \nn
\end{split}
\ee 
{\rm (iv):} For each $(i, p)$, $(Z_n^{i,p})_{n=0}^N$ is a Markov process i.e.,  $\mbb{E}^{i,p}[f(Z_n^{i,p})|\calf_{t_k}^{i,p}]=\ex^{i,p}[f(Z_n^{i,p})|Z_k^{i,p}]$
for every bounded measurable function $f$  on $\calz_n^p$ and $k\leq n$.\\
{\rm (v):} $(Y_n)_{n=0}^N$ is a Markov process i.e.,  $\mbb{E}^0[f(Y_n)|\calf_{t_k}^0]=\ex^0[f(Y_n)|Y_k]$
for every bounded measurable function $f$ on $\caly_n$ and $k\leq n$. \\
{\rm (vi):} The transition probabilities of $(S_n)_{n=0}^N$ satisfy, for every $0\leq n\leq N-1$, a.s.,
\be
\begin{split}
\mbb{P}^0(S_{n+1}=\wt{u}S_{n}|\calf_{t_n}^0)&=\mbb{P}^0(S_{n+1}=\wt{u}S_n|\bS^n,Y_n)=:p_{n}(\bS^{n},Y_{n}), \\
\mbb{P}^0(S_{n+1}=\wt{d}S_{n}|\calf_{t_n}^0)&=\mbb{P}^0(S_{n+1}=\wt{d}S_n|\bS^n,Y_n)=:q_{n}(\bS^{n},Y_{n}), \nn
\end{split}
\ee
where $p_{n}, q_n~(:=1-p_n):\cals^{n}\times \caly_n\rightarrow \mbb{R}, ~0\leq n\leq N-1$ are bounded measurable functions
satisfying 
\be
0<p_n(\bs,y), q_n(\bs,y)<1 \nn
\ee
for every $(\bs,y)\in \cals^n\times \caly_n$.
\end{assumption}

We also introduce the terminal liabilities $F^p$ and the incremental endowments $(g_n^p)_{n=1}^N$,
which may have different functional forms across the populations $p=1,\ldots, m$ to capture the diversity of financial entities.

\begin{assumption}
\label{assumption-Rm2} For each $1\leq p\leq m$, \\
{\rm (i):} the function $F^p:\cals^N\times \caly_N\times \calz_N^p \rightarrow \mbb{R}$ is measurable and bounded.\\
{\rm (ii):} for every $1\leq n\leq N$, the function $g_n^p:\cals^n\times\caly_n\times \calz_n^p\rightarrow \mbb{R}$
is measurable and bounded.\\
{\rm (iii):} every agent is a price-taker in the sense that they consider the stock price process (and hence its transition 
probabilities specified in Assumption~\ref{assumption-Rm1} (vi)) to be exogenously given  by the collective actions 
of the others and unaffected by the agent's own trading strategies.
\end{assumption}

We suppose that each agent-$(i,p)$ solves the optimization problem
\be
\sup_{(\phi_n^{i,p}, c_n^{i,p})_{n=0}^{N-1}\in \mbb{A}^{i,p}} U_0^{i,p}, \nn
\ee
over the admissible space defined by
\be
\mbb{A}^{i,p}:=\bigl\{(\phi_n^{i,p}, c_n^{i,p})_{n=0}^{N-1}: (\phi_n^i, c_n^i) \text{ is an $\calf_n^{0,(i,p)}$-measurable $\mbb{R}^2$-valued random variable} \bigr\}. \nn
\ee
The $(\calf_{t_n}^{0,(i,p)})_{n=0}^N$-adapted process of the recursive utilities $(U^{i,p}_n)_{n=0}^N$ is defined by
\be
U_n^{i,p}:=-\frac{1}{\zeta_i^p}\log\Bigl\{\exp({-\zeta_i^p c_n^{i,p}})\Del+\del_i^p\exp\Bigl(\frac{\psi_i^p}{\gamma_i^p}
\log\bigl(\ex^{0,(i,p)}[e^{-\gamma_i^p U_{n+1}^{i,p}}|\calf_{t_n}^{0,(i,p)}]\bigr)\Bigr)\Bigr\} \nn
\ee
with the terminal condition
\be
U_N^{i,p}:=X_N^{i,p}-F^p(\bS^N, Y_N, Z_N^{i,p}). \nn
\ee
Here, $(X_n^{i,p})_{n=0}^N$ denotes the wealth process of agent-$(i,p)$ which follows
\be
X_{n+1}^{i,p}=\beta(X_n^{i,p}-c_n^{i,p}\Del)+\phi_n^{i,p}R_{n+1}+g_{n+1}^p(\bS^{n+1},Y_{n+1},Z_{n+1}^{i,p}), \nn
\ee
where $X_0^{i,p}=\xi_i^p$. The interpretations of the coefficients are the same as in Section~\ref{sec-recursive}.

Under Assumptions~\ref{assumption-Rm1} and \ref{assumption-Rm2}, 
a direct application of Theorem~\ref{th-R1} for each population $p=1,\ldots, m$ provides the 
optimal solution $(\phi^{i,p,*}_{n-1}, c^{i,p,*}_{n-1})_{n=1}^N$. In particular, we have
$\phi^{i,p,*}_{n-1}:=\phi^{i,p,*}_{n-1}(\bS^{n-1},Y_{n-1},Z_{n-1}^{i,p},\vr_i^p)$, 
where for each $(\bs,y,z^{i,p},\vr_i^p)\in \cals^{n-1}\times \caly_{n-1}\times \calz_{n-1}^p\times \Gamma^p$, 
\be
\label{phi-m-optimal}
\phi^{i,p,*}_{n-1}(\bs,y,z^{i,p},\vr_i^p):=\frac{1}{\gamma_i^p \eta_n^{i,p}(u-d)}
\Bigl\{ \log\Bigl(-\frac{p_{n-1}(\bs,y)u}{q_{n-1}(\bs,y)d}\Bigr)+\log\bigl(f_{n-1}^p(\bs, y, z^{i,p},\vr_i^p)\bigr)\Bigr\}. \
\ee
Here,  $f_{n-1}^p$ is given by  $(\ref{th-R1-fnm1})$ by replacing $F$ with $F^p$ and $g_n$ with $g_n^p$
in its recursive definition. Similarly, $\eta_n^{i,p}$ is given by $(\ref{th-R1-etanm1})$ with $\psi_i\rightarrow \psi_i^p$ and 
$\zeta_i\rightarrow \zeta_i^p$.

We now consider the large population limit and the associated market-clearing condition. As  discussed in \cite{Fujii,Fujii-CP}, we need to keep
track of the ratio of population size. Let us denote the number of agents in population $p$ by $N_p$
and set $\caln:=N_1+\cdots +N_m$. We use $w_p:=N_p/\caln $ to denote the relative size of population $p$.
We obviously have the following decomposition:
\be
\label{mp-decomp}
\begin{split}
\frac{1}{\caln}\sum_{p=1}^m \sum_{i=1}^{N_p} \phi^{i,p, *}_{n-1}=\sum_{p=1}^m w_p \Bigl(\frac{1}{N_p}\sum_{i=1}^{N_p} \phi^{i,p, *}_{n-1}\Bigr). 
\end{split}
\ee
We thus consider the limit $\caln\rightarrow \infty$ while keeping $w_p, 1\leq p\leq m$ constant,
and then introduce the following concept of mean-field equilibrium in the presence of an external order flow:
\begin{definition}
\label{def-multi}
We say that the system is in the mean-field equilibrium if
\be
 \lim_{\caln\rightarrow\infty}\frac{1}{\caln}\sum_{p=1}^m \sum_{i=1}^{N_p} \phi^{i,p, *}_{n-1}(\bS^{n-1},Y_{n-1},Z_{n-1}^{i,p},\vr_i^p)=L_{n-1}(\bS^{n-1},Y_{n-1}), \nn
\ee
$\mbb{P}$-a.s. for every $1\leq n\leq N$ with $\phi^{i,p, *}_{n-1}$ defined by $(\ref{phi-m-optimal})$,
where the large population limit is taken with the population ratios $(w_p)_{p=1}^m$ kept constant.
\end{definition}

To simplify the notation and facilitate the subsequent discussion, we introduce the following effective variables:
\be
\begin{split}
&\calt_n^{i,p}:=\frac{1}{\gamma_i^p \eta_n^{i,p}}, \qquad \calt_n:=\sum_{p=1}^m w_p\ex^{1,p}\Bigl[\frac{1}{\gamma_1^p \eta_n^{1,p}}\Bigr],  \\
&\calv_{n-1}^{i,p}(\bs,y,z^{i,p},\vr_i^p):=\frac{\log f_{n-1}^p(\bs,y,z^{i,p},\vr_i^p)}{\gamma_i^p\eta_n^{i,p}}, \\
&\calv_{n-1}(\bs,y):=\sum_{p=1}^m w_p \ex^{1,p}\Bigl[\frac{\log f_{n-1}^p(\bs,y,Z_{n-1}^{1,p},\vr_1^p)}{\gamma_1^p \eta_n^{1,p}}\Bigr].\nn
\end{split}
\ee
Here, $\calt^{i,p}_n$, which is $\calf^{i,p}_0$-measurable, represents the \textit{effective risk tolerance} of agent-$(i,p)$ at $t_n$. 
Note that $\calt_n$ corresponds to the \textit{aggregated risk tolerance} of the market, satisfying the relation:
\be
\label{effective-conv-1}
\calt_n=\sum_{p=1}^m w_p\ex^{1,p}[\calt_n^{1,p}].
\ee
Similarly, $\calv_{n-1}^{i,p}:\cals^{n-1}\times \caly_{n-1}\times \calz_{n-1}^p\times \Gamma^p\rightarrow \mbb{R}$
is a  measurable function representing the sensitivity of the effective liability at $t_{n-1}$ for agent-$(i,p)$,
where the dependence on $\gamma_i^p\eta_n^{i,p}$ is incorporated by the argument $\vr_i^p$.
Thus, strictly speaking, the superscript $i$ in $\calv^{i,p}_{n-1}$ is redundant, but we retain it to 
clearly associate the variable with the relevant agent.
$\calv_{n-1}:\cals^{n-1}\times \caly_{n-1}\rightarrow \mbb{R}$ denotes the aggregated sensitivity of the effective liability, which satisfies
\be
\label{effective-conv-2}
\calv_{n-1}(\bs,y)=\sum_{p=1}^m w_p\ex^{1,p}\bigl[\calv_{n-1}^{1,p}(\bs,y,Z_{n-1}^{1,p},\vr_1^p)\bigr]. 
\ee
With these effective variables, the optimal control $(\ref{phi-m-optimal})$ can be expressed as
\be
\phi^{(i,p),*}_{n-1}(\bs,y,z^{i,p},\vr_i^p)=\frac{1}{u-d}\Bigl\{
\log\Bigl(-\frac{p_{n-1}(\bs,y)u}{q_{n-1}(\bs,y)d}\Bigr)\calt_n^{i,p}+\calv_{n-1}^{i,p}(\bs,y,z^{i,p},\vr_i^p)\Bigr\}.  \nn
\ee

We can now derive the equilibrium transition probabilities among the $m$ populations.
\begin{theorem}
\label{th-multi}
Let Assumptions~\ref{assumption-Rm1}, \ref{assumption-Rm2} and \ref{assumption-R3} be in force.
Then there exists a unique mean-field equilibrium. The associated transition probabilities of the stock price are given by
\be
\begin{split}
&p_{n-1}(\bs,y):=\mbb{P}^0\Bigl(S_n=\wt{u}S_{n-1}|(\bS^{n-1},Y_{n-1})=(\bs,y)\Bigr)\\
&=(-d)\Big/ \left\{ u\exp\left(\frac{\calv_{n-1}(\bs,y)-(u-d)L_{n-1}(\bs,y)}{\calt_n}
\right)-d\right\} \nn
\end{split}
\ee
for every $(\bs,y)\in \cals^{n-1}\times \caly_{n-1}$, $1\leq n\leq N$. 
Here, for each $p=1,\ldots, m$, the functions $f_{n-1}^p:\cals^{n-1}\times \caly_{n-1}\times \calz_{n-1}^p\times \Gamma^p\rightarrow \mbb{R}$,  $1\leq n\leq N$, which are constitutive components of $(\calv_{n-1})_{n=1}^N$, are 
determined by the backward induction in Theorem~\ref{th-R1}
applied to each population $p$, with the transition probabilities replaced by those given above at each step.
Under the above transition probabilities,
the optimal strategy of agent-$(i,p)$ is given by, for each $(\bs,y,z^{i,p},\vr_i^p)\in \cals^{n-1}\times\caly_{n-1}\times \calz_{n-1}^p\times \Gamma^p$,
\be
\label{multi-mfe-optimal}
\phi^{(i,p),*}_{n-1}(\bs,y,z^{i,p},\vr_i^p)=\frac{\calt_n^{i,p}}{\calt_n}L_{n-1}(\bs,y)+
\frac{1}{u-d}\Bigl(\calv_{n-1}^{i,p}(\bs,y,z^{i,p},\vr_i^p)-\frac{\calt_n^{i,p}}{\calt_n}\calv_{n-1}(\bs,y)\Bigr). 
\ee
Moreover, there exists some positive constant $\calc_{n-1}$ such that
\be
\ex\Bigl|\frac{1}{\caln}\sum_{p=1}^m \sum_{i=1}^{N_p}\phi^{i,p,*}_{n-1}(\bS^{n-1},Y_{n-1},Z_{n-1}^{i,p},\vr_i^p)-L_{n-1}(\bS^{n-1},Y_{n-1})\Bigr|^2\leq
\frac{\calc_{n-1}}{\caln} \nn
\ee
for every $1\leq n\leq N$, which gives the convergence rate in the large population limit.
\end{theorem}
\begin{proof}
See Appendix~\ref{A-th-multi}.
\end{proof}

\begin{remark}[Economic Interpretation of the Equilibrium Strategy]
The expression $(\ref{multi-mfe-optimal})$ provides a clear economic interpretation of the equilibrium strategy. 
It can be decomposed into two distinct components:
\be
\phi^{(i,p),*}_{n-1} = \underbrace{\frac{\calt_n^{i,p}}{\calt_n} L_{n-1}}_{\text{(I) Supply Distribution}} 
+ \underbrace{\frac{1}{u-d}\left( \calv_{n-1}^{i,p}- \frac{\calt_n^{i,p}}{\calt_n}\calv_{n-1} \right)}_{\text{(II) Hedge Distribution}}. \nn
\ee
\begin{itemize}[noitemsep]
\item[(I)] The first term represents the sharing of the external supply $L_{n-1}$. 
Each agent absorbs a portion of the supply proportional to their effective risk tolerance $\calt_n^{i,p}$ 
relative to the market aggregate $\calt_n$. 
\item[(II)] The second term corresponds to the hedging demand against the effective liability at $t_n$. 
Since $\calv_{n-1}^{i,p}$ represents the sensitivity of the liability (scaled by risk aversion), 
the term $\calv_{n-1}^{i,p}/(u-d)$ essentially corresponds to the Delta hedge required for agent-$(i,p)$. 
However, since the market must clear, agents cannot simply hold their desired hedge. 
Instead, the aggregate hedging demand of the market, $\calv_{n-1}/(u-d)$, is redistributed back to the agents 
according to their risk tolerance shares. 
\end{itemize}
The above decomposition for the equilibrium strategy is universal. One can easily confirm
that the same decomposition holds in Theorems~\ref{th-t2}, \ref{th-R2}, and also Theorem~\ref{th-R2-P} in the next section.   
\end{remark}

As discussed in Remark~\ref{remark-path}, 
if we turn off the path-dependence in $F^p$ and $(g_n^p)$,  then 
the simpler form of transition probabilities, $(p_{n-1}(s,y),q_{n-1}(s,y)), ~(s,y)\in \cals_{n-1}\times \caly_{n-1}$, is sufficient to clear the market.
In this case, at least, numerical costs would not be so high and
one can calculate the equilibrium price distribution in the same way as in the next Section~\ref{sec-numerical}.
We can even mix the populations
with standard exponential utilities and the recursive utilities in a similar manner.

It should be noted that the analytical tractability achieved in this section is quite
remarkable when compared with the situation in continuous-time settings.
In fact, attempting to  solve 
the corresponding problem in the formulation of Fujii \& Sekine~\cite{Fujii-Sekine1} would lead to 
a coupled system of mean-field qg-BSDEs. Also in the optimal storage-type modeling 
adopted in many of the existing literature, such as \cite{Fujii, Giulia, Shrivats}, the multi-population extension would produce a coupled system of FBSDEs of  conditional McKean-Vlasov type. The well-posedness  of these equations is substantially more challenging than in the 
single population case, not to mention its numerical evaluation.

\section{Subjective Measures: Stochastic Bias and Equilibrium}
\label{sec-subjective}
This section addresses the question of whether every agent must adopt the correct objective probability measure
$\mbb{P}^0$ for the dynamics of the stock price to achieve the market-clearing equilibrium. 
All existing literature on MFG implicitly adopts the rational expectation hypothesis, assuming that every agent shares correct knowledge of the objective probability distributions governing the relevant state processes or outcomes.
A recent paper by Moll \& Ryzhik~\cite{Moll-irrational} (2025) studies MFGs without rational expectations.
They show that, in some cases, departing from rational expectations completely sidesteps the Master equation, 
allowing for the solution of much simpler finite-dimensional HJB equations instead.

In the preceding sections, we adopt Assumptions~\ref{assumption-T2} (ii), \ref{assumption-R2} (iii), and \ref{assumption-Rm2} (iii),
all of which assume that every agent-$i$ possesses true knowledge of the stock dynamics by sharing the objective measure 
$\mbb{P}^0$. Specifically, they use the objective probability measure $\mbb{P}^{0,i}:=\mbb{P}^0\otimes\mbb{P}^i$,
which incorporates the correct objective distributions for $(S,Y)$ and the agents' idiosyncratic shocks.
As emphasized in \cite{Moll-irrational}, it is more natural to suppose that each agent-$i$ adopts a subjective measure,
which is estimated, for example, by a statistical technique or  based on a plausible economic model.
Then the resultant dynamics of the stock price under  the subjective  measure  is likely  to exhibit 
a bias relative to the true dynamics. Moreover, this bias is expected to change stochastically dependent on the agents' risk characteristics 
as well as the market environment. 
The ultimate goal of this section is to determine the objective transition probability of the stock price under the measure
$\mbb{P}^0$ that clears the market among these agents operating under their respective subjective measures.

\subsection{The setup and notation}
In this section,  we directly extend the results in Section~\ref{sec-recursive} to incorporate agents' subjective measures on the stock dynamics. 
We adopt the same setup and notation
as in Section~\ref{sec-setup-recursive}. The key novelty in this section is captured by the following assumption:
\begin{assumption}
\label{assumption-R-P}
Each agent-$i$ adopts a subjective probability measure $\calp^{0,i}$ on $(\Omega^{0,i},\calf^{0,i}, (\calf^{0,i}_{t_n})_{n=0}^N)$
which may differ from the objective measure $\mbb{P}^{0,i}$. The two measures $\calp^{0,i}$ and $\mbb{P}^{0,i}$
coincide for the dynamics of the variables $(Y,Z^i,\xi_i,\vr_i)$ but differ only in the stock dynamics in the following way:
\be
\begin{split}
\calp^{0,i}(S_{n+1}=\wt{u}S_n|\calf_{t_n}^{0,i})&=\calp^{0,i}(S_{n+1}=\wt{u}S_n|\bS^n,Y_n, Z_n^i,\vr_i)=:\mdp_n(\bS^n,Y_n,Z_n^i,\vr_i), \\
\calp^{0,i}(S_{n+1}=\wt{d}S_n|\calf_{t_n}^{0,i})&=\calp^{0,i}(S_{n+1}=\wt{d}S_n|\bS^n,Y_n,Z_n^i,\vr_i)=:\mdq_n(\bS^n,Y_n,Z_n^i,\vr_i), \nn
\end{split}
\ee
where $\mdp_n$ and $\mdq_n:=1-\mdp_n$ are measurable functions defined on $\cals^n\times\caly_n\times\calz_n\times\Gamma$
and satisfy, for each $n=0,\ldots, N-1$, 
\be
\label{sub-transition-formula}
\frac{\mdp_n(\bs ,y,z^i,\vr_i)}{\mdq_n(\bs,y,z^i,\vr_i)}=\varpi_n(\bs,y,z^i,\vr_i)\Bigl(\frac{p_n(\bs,y)}{q_n(\bs,y)}\Bigr), 
\ee
where $p_{n}, q_{n}$ are the transition probabilities under the objective measure given in Assumption~\ref{assumption-R1} (vi) and
$\varpi_n: \cals^n\times\caly_n\times \calz_n\times\Gamma\rightarrow \mbb{R}$ is a measurable function that expresses
the agents' biased view on the transition probabilities.
There exist some positive constants $0<\ul{\varpi}<\ol{\varpi}<\infty$ such that
\be
\ul{\varpi}\leq \varpi_n(\bs,y,z^i,\vr_i)\leq \ol{\varpi} \nn
\ee
for every $(\bs,y,z^i,\vr_i)\in \cals^n\times\caly_n\times\calz_n\times\Gamma$, $0\leq n\leq N-1$.
\end{assumption}
\noindent
We denote by $\ex_\calp^{0,i}[\cdot]$ the expectation with respect to $\calp^{0,i}$. Note that $\calp^{0,i}$ 
is equivalent to $\mbb{P}^{0,i}$ due to the strict positivity and boundedness of $(\varpi_n)_n$.
This equivalence guarantees that each agent's subjective view remains arbitrage-free, 
which is a fundamental requirement 
for the well-posedness and economic consistency of the individual optimization problem.

\bigskip
In this model, the individual bias on the transition probabilities at $t_n$ is captured by the function $\varpi_n$. 
Specifically, $\varpi_n>1$ (resp., $\varpi_n<1$)
implies a positive (resp., negative) view on the stock performance for the period $[t_n,t_{n+1}]$, which is conditional on
the time $t_n$ realizations of  the stock price history,  macroeconomic/environmental factors $Y$, 
idiosyncratic shocks $Z^i$, as well as the agent's risk coefficients.
For example, if agent-$i$ is relatively more cautious (resp., optimistic), they might have $\varpi_n<1$ (resp., $\varpi_n>1$) 
when the stock price and the macroeconomic factor take large (i.e., good) values.

\subsection{The individual optimization problem}
As in Section~\ref{sec-rec-individual}, the $(\calf_{t_n}^{0,i})_{n=0}^N$-adapted process of utilities $(U_n^i)_{n=0}^N$ 
is recursively defined by
\be
\label{def-RU-P}
U_n^i:=-\frac{1}{\zeta_i}\log\Bigl\{\exp(-\zeta_i c_n^i)\Del+\del_i \exp\Bigl(\frac{\psi_i}{\gamma_i}\log \bigl(\ex^{0,i}_\calp[e^{-\gamma_i U_{n+1}^i}
|\calf_{t_n}^{0,i}]\bigr)\Bigr), 
\ee
with the terminal condition $U_N^i:=X_N^i-F(\bS^N,Y_N,Z_N^i)$.
The sole difference from $(\ref{def-RU})$ is the use of the expectation $\ex^{0,i}_{\calp}[\cdot]$ with respect to the subjective measure $\calp^{0,i}$.
The wealth process of agent-$i$, $(X_n^i)_{n=0}^N$, follows the same dynamics as in $(\ref{eq-rec-wealth})$,
and the agent problem is also given by  $(\ref{problem-R1})$ with the identical admissible space $\mbb{A}^i$.

\begin{theorem}
\label{th-R1P}
Let Assumptions~\ref{assumption-R1}, \ref{assumption-R2}~\text{\rm (i), (ii)}, and \ref{assumption-R-P} be in force.
Then the problem $(\ref{problem-R1})$ with utility defined by $(\ref{def-RU-P})$ 
has an a.s. unique optimal solution $(\phi_{n-1}^{i,*}, c_{n-1}^{i,*})_{n=1}^{N}$,
where $(\phi_{n-1}^{i,*})_{n=1}^N$ and $(c_{n-1}^{i,*})_{n=1}^N$ are  bounded processes defined by
measurable functions $\phi_{n-1}^{i,*}:\cals^{n-1}\times \caly_{n-1}\times \calz_{n-1}\times \Gamma\rightarrow \mbb{R}$
and $c_{n-1}^{i,*}:\mbb{R}\times \cals^{n-1}\times \caly_{n-1}\times \calz_{n-1}\times \Gamma\rightarrow \mbb{R}$
such that $\phi_{n-1}^{i,*}:=\phi^{i,*}_{n-1}(\bS^{n-1},Y_{n-1},Z_{n-1}^i,\vr_i)$
and $c_{n-1}^{i,*}:=c^{i,*}_{n-1}(X_{n-1}^i, \bS^{n-1}, Y_{n-1}, Z_{n-1}^i, \vr_i)$ respectively.
Furthermore, all the expressions $(\ref{th-R1-phinm1})$-$(\ref{th-R1-etanm1})$ and their properties
stated in Theorem~\ref{th-R1} hold, provided that the objective transition 
probabilities $(p_{n-1}(\bs,y), q_{n-1}(\bs,y))$ in $(\ref{th-R1-phinm1})$ and $(\ref{th-R1-Vtildenm1})$ are replaced by the subjective probabilities 
$(\mdp_{n-1}(\bs,y,z^i,\vr_i)$, $\mdq_{n-1}(\bs,y,z^i,\vr_i))$ for each $(\bs, y, z^i,\vr_i)\in \cals^{n-1}\times \caly_{n-1}\times \calz_{n-1}\times \Gamma$.
\end{theorem}
\begin{proof}
See Appendix~\ref{A-th-R1P}.
\end{proof}

From Theorem~\ref{th-R1P}, we obtain, for each $(\bs,y,z^i,\vr_i)\in \cals^{n-1}\times \caly_{n-1}\times \calz_{n-1}\times \Gamma$,
\be
\begin{split}
\phi_{n-1}^{i,*}(\bs,y, z^i,\vr_i)=\frac{1}{\gamma_i \eta_n^i (u-d)}\Bigl\{ \log\Bigl(-\frac{\mdp_{n-1}(\bs,y,z^i,\vr_i)u}{\mdq_{n-1}(\bs,y,z^i,\vr_i)d}\Bigr)
+\log\bigl(f_{n-1}(\bs, y, z^i,\vr_i)\bigr)\Bigr\}, \nn
\end{split}
\ee
which is the optimal strategy of agent-$i$. Using Assumption~\ref{assumption-R-P}, we have
\be
\label{rec-phinm1-P}
\phi_{n-1}^{i,*}(\bs,y, z^i,\vr_i)=\frac{1}{\gamma_i \eta_n^i (u-d)}\Bigl\{ \log\Bigl(-\frac{p_{n-1}(\bs,y)u}{q_{n-1}(\bs,y)d}\Bigr)
+\log\bigl(f^\pi_{n-1}(\bs, y, z^i,\vr_i)\bigr)\Bigr\}, 
\ee
where $f_{n-1}^\pi:\cals^{n-1}\times \caly_{n-1}\times \calz_{n-1}\times \Gamma\rightarrow \mbb{R}$
is a strictly positive, bounded, measurable function defined by
\be
\label{f-subjective}
f^\pi_{n-1}(\bs, y, z^i,\vr_i):=\varpi_{n-1}(\bs,y,z^i,\vr_i)f_{n-1}(\bs,y,z^i,\vr_i). 
\ee
This suggests that the stochastically biased views manifest effectively  as a modification to the effective liabilities 
and/or incremental endowments.~\footnote{In addition to the multiplicative factors $(\varpi_n)_n$, the subjective transition probabilities  $(\mdp_n, \mdq_n)_n$ also change the values of 
$(V_n, \wt{V}_n)_n$ and consequently of $(f_n)_n$ through their updates in the recursive formula. }

\subsection{Mean-field equilibrium among the agents with subjective measures}
\label{sec-mfe-subjective}
As a main goal of this section, we shall now derive a set of transition probabilities $(p_{n-1}, q_{n-1})$ of the stock price
under the objective measure $\mbb{P}^0$ so that the mean-field equilibrium holds among the agents operating under 
respective subjective measures with stochastically biased estimates on the stock transition probabilities.
The mean-field market-clearing condition is defined in Definition~\ref{def-mfe-rec}
with $(\phi^{i,*}_{n-1})$ given by Theorem~\ref{th-R1P}. This is still characterized by equation $(\ref{rec-clearing})$.
Using the expression $(\ref{rec-phinm1-P})$, it is straightforward to modify Theorem~\ref{th-R2}
for the current setting:
\begin{theorem}
\label{th-R2-P}
Let Assumptions~\ref{assumption-R1}, \ref{assumption-R2}~\text{\rm (i), (ii)}, \ref{assumption-R3} and \ref{assumption-R-P} be in force.
Then there exists a unique mean-field equilibrium. The associated objective transition probabilities of the stock price are given by
\be
\label{mfe-transition-P}
\begin{split}
&p_{n-1}(\bs,y):=\mbb{P}^0\Bigl(S_n=\wt{u}S_{n-1}|(\bS^{n-1},Y_{n-1})=(\bs,y)\Bigr)\\
&=(-d)\Big/ \Bigl\{ u\exp\Bigl(\frac{1}{\ex^1[1/(\gamma_1\eta_n^1)]}\Bigl[
\ex^1\Bigl(\frac{\log (f^\pi_{n-1}(\bs,y,Z^1_{n-1},\vr_1))}{\gamma_1\eta_n^1}\Bigr)-(u-d)L_{n-1}(\bs,y)\Bigr]\Bigr)-d\Bigr\} 
\end{split}
\ee
for every $(\bs,y)\in \cals^{n-1}\times \caly_{n-1}$, $1\leq n\leq N$. 
Here, $f_{n-1}^\pi:=\varpi_{n-1}\times f_{n-1}$ is defined by $(\ref{f-subjective})$ and the other variables are as defined in  Theorem~\ref{th-R1P}.
Specifically, $f_{n-1}:\cals^{n-1}\times\caly_{n-1}\times \calz_{n-1}\times \Gamma\rightarrow \mbb{R}$, $1\leq n\leq N$,
are given by the backward induction in Theorem~\ref{th-R1},
with the objective transition probabilities replaced by the subjective ones,  
which are determined via  $(\ref{sub-transition-formula})$
using the bias functions $(\varpi_n)$ and the objective transition probabilities given above at each step. 
Under the above transition probabilities, the optimal strategy of agent-$i$ is given by
\be
\begin{split}
\phi^{i,*}_{n-1}(\bs,y,z^i,\vr_i)&=\frac{1}{(u-d)}\Bigl\{\frac{\log f^\pi_{n-1}(\bs,y,z^i,\vr_i)}{\gamma_i \eta_n^i}
-\frac{1/(\gamma_i \eta_n^i)}{\ex^1[1/(\gamma_1\eta_n^1)]}\ex^1\Bigl(\frac{\log f^\pi_{n-1}(\bs,y,Z_{n-1}^1,\vr_1)}{\gamma_1\eta_n^1}\Bigr)
\Bigr\}\\
&+\frac{1/(\gamma_i \eta_n^i)}{\ex^1[1/(\gamma_1\eta_n^1)]}L_{n-1}(\bs,y). \nn
\end{split}
\ee
Moreover, there exists some positive constant $\calc_{n-1}$ such that
\be
\ex\Bigl|\frac{1}{N_p}\sum_{i=1}^{N_p}\phi^{i,*}_{n-1}(\bS^{n-1},Y_{n-1},Z_{n-1}^i,\vr_i)-L_{n-1}(\bS^{n-1},Y_{n-1})\Bigr|^2\leq
\frac{\calc_{n-1}}{N_p} \nn
\ee
for every $1\leq n\leq N$, which gives the convergence rate in the large population limit.
\end{theorem}
\begin{proof}
See Appendix~\ref{A-th-R2-P}.
\end{proof}

It is instructive to decompose $f^\pi$ in the expression $(\ref{mfe-transition-P})$ in the following way:
\be
\begin{split}
&p_{n-1}(\bs,y):=\mbb{P}^0\Bigl(S_n=\wt{u}S_{n-1}|(\bS^{n-1},Y_{n-1})=(\bs,y)\Bigr)\\
&=(-d)\Big/ \left\{ u\exp\left(\frac{\ex^1\Bigl[\frac{\log f_{n-1}(\bs,y,Z^1_{n-1},\vr_1)}{\gamma_1\eta_n^1}\Bigr]
-(u-d)L_{n-1}(\bs,y)+\ex^1\Bigl[\frac{\log \varpi_{n-1}(\bs,y,Z^1_{n-1},\vr_1)}{\gamma_1\eta_n^1}\Bigr] }{\ex^1\Bigl[\frac{1}{\gamma_1\eta_n^1}
\Bigr]}\right)-d\right\}. \nn
\end{split}
\ee

This result implies that, apart from the subtle effects induced by the changes in $f_{n-1}$, 
the aggregated effect of the biased views on the stock price transition probabilities
can also manifest  as an external order flow term $L_{n-1}$ in the standard rational expectation setting.
Specifically, it is clear that the negative view $(\varpi_{n-1}<1)$ adds to the external supply $L_{n-1}$, 
and the positive one $(\varpi_{n-1}>1)$ does the opposite.
Combined with the discussions in Section~\ref{sec-implications},  
the above result implies that the existence of a large portion of agents whose bias
reflects a counter-cyclical,  contrarian view---i.e., exhibiting a cautious (pessimistic) view $(\varpi_{n-1}<1)$ when the 
stock price and/or market-wide economic factor are at high levels, and, conversely, an optimistic 
view $(\varpi_{n-1}>1)$ when they are at low levels---tends to make the equilibrium price distributions more fat-tailed.

\begin{remark}[Path Dependence and Extension to Multiple Populations]
We remark that the discussions in Remark~\ref{remark-path} regarding the necessary length of the price trajectory in the transition probabilities
remain valid for the results in this section.
Moreover, the framework of multi-population equilibrium presented in Section~\ref{sec-multi-p} can be readily extended
to the current model for agents with subjective measures.
Specifically, we only need to replace, for each $p=1,\ldots, m$,  the set of functions $(f^p_{n})$ by $(f^{\pi, p}_n)$
where $f^{\pi,p}_n=\varpi^p_n \times f^p_n$ with population-specific bias functions $(\varpi_n^p)$.
\end{remark}

\section{Numerical examples and implications}
\label{sec-numerical}
In this section, we provide some numerical examples for the models introduced in Sections~\ref{sec-terminal} and \ref{sec-recursive}, focusing on cases without path dependence. We also provide an example with stochastically biased agents developed in Section~\ref{sec-subjective}.
Since the models are too flexible for a thorough analysis, we focus only on a few simple setups
to illustrate characteristic behaviors of the mean-field price distributions.
In order to reduce computational cost, we assume that $Y$ and $Z^i$ are  one-dimensional discrete processes taking values on binomial trees,
and that all $\calf^i_0$-measurable random variables are uniformly distributed over finite sets.

For our analytical formulation, the finite state space for $(S,Y)$ has played an important role.
However, there is no  requirement for $Y$ to follow a binomial process.
Moreover, the coefficients $\vr_i:=(\gamma_i,\zeta_i, \del_i,\psi_i)$
and the process $(Z^i_n)$ can have continuous distributions. The specific assumptions made in this section 
are solely for numerical convenience.

\subsection{Utility with terminal liability}
\label{sec-num-terminal}
We first consider the model discussed in Section~\ref{sec-terminal}.
$\gamma_i$ is assumed to be uniformly distributed over the $(N_\gamma+1)$ discrete values given by
\be
\gamma_i(k_\gamma):=\ul{\gamma}+(\ol{\gamma}-\ul{\gamma})k_\gamma/N_\gamma, \quad k_\gamma=0, \cdots, N_\gamma. \nn
\ee
The process $(Z_n^i)_{n=0}^N$ is supposed to follow a one-dimensional binomial process modeled by
\be 
Z_{n+1}^i=Z_n^i R_{n+1}^i, \nn
\ee
where $(R_{n}^i)$ is an $(\calf^i_{t_n})$-adapted process taking values either $u_z$ or $d_z$.
Specifically,  $R_{n}^i=u_z$ occurs with probability $p_z$ and $R_{n}^i=d_z$ with $q_z:=1-p_z$.
We take $u_z=(d_z)^{-1}=\exp(\sigma_z \sqrt{\Del})$. We also assume $Z_0^i=z_0 \in (0,\infty)$ is common for all the agents in order to reduce
computational costs. We model the process $(Y_n)_{n=0}^N$ similarly but it is assumed to follow an approximate Gaussian process:
\be
Y_{n+1}=Y_n +R_{n+1}^y, \nn
\ee 
where $(R_n^y)$ is an $(\calf^0_{t_n})$-adapted process taking values either $u_y$ or $d_y$, 
where $R_{n}^y=u_y$ with probability $p_y$ and $R_{n}^y=d_y$ with $q_y:=1-p_y$.
We take $u_y=(-d_y)=\sigma_y \sqrt{\Del}$.
Finally, for the stock-price process $(S_n)$, we set $\wt{u}=(\wt{d})^{-1}=\exp(\sigma \sqrt{\Del})$ and $S_0=1.0$.
The parameter values to be used throughout this subsection are summarized in Table~\ref{tab-param-1} below. 
We recall that the initial wealth $\xi_i$
is irrelevant for our analysis.
\begin{table}[h]
    \footnotesize
    \centering
    \begin{tabular}{c c c c c c c c c c c c c c c } 
        \toprule
        \text{parameter} &  $\ul{\gamma}$ &  $\ol{\gamma}$ &  $N_\gamma$ & $z_0$ & $\sigma_z$ & $p_z$ &  $Y_0$ & $\sigma_y$ & $p_y$ 
		& $S_0$ & $\sigma$  & $r$ & $T$ & $N$ \\
        \midrule
	  \text{value} & 0.5 & 1.5 & 4  & 1.0 & 12\% & 0.5 & 1.0 & 12\% & 0.5 &1.0 & 15\%&  3.3\% & 3yr &48\\
        \bottomrule
    \end{tabular}
    \caption{ parameter values }
  \label{tab-param-1}
\end{table}

\begin{figure}[H]
\vspace{0mm}
    \centering
    \begin{minipage}[t]{0.49\textwidth}
        \centering
        \includegraphics[width=\linewidth]{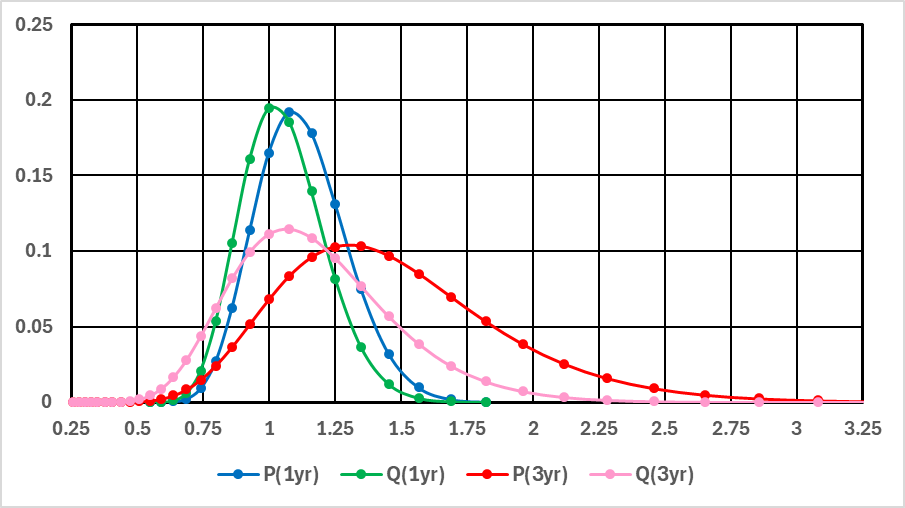}
	\caption{\footnotesize Comparison of the marginal price distributions under the equilibrium measure $(\mbb{P})$ and the risk-neutral measures $(\mbb{Q})$
	at 1-year and 3-year points for $(\ref{Terminal-F1})$.}
	 \label{fig-marginal-1}
    \end{minipage}%
    \hspace{0.01\textwidth} 
    \begin{minipage}[t]{0.49\textwidth}
        \centering
        \includegraphics[width=\linewidth]{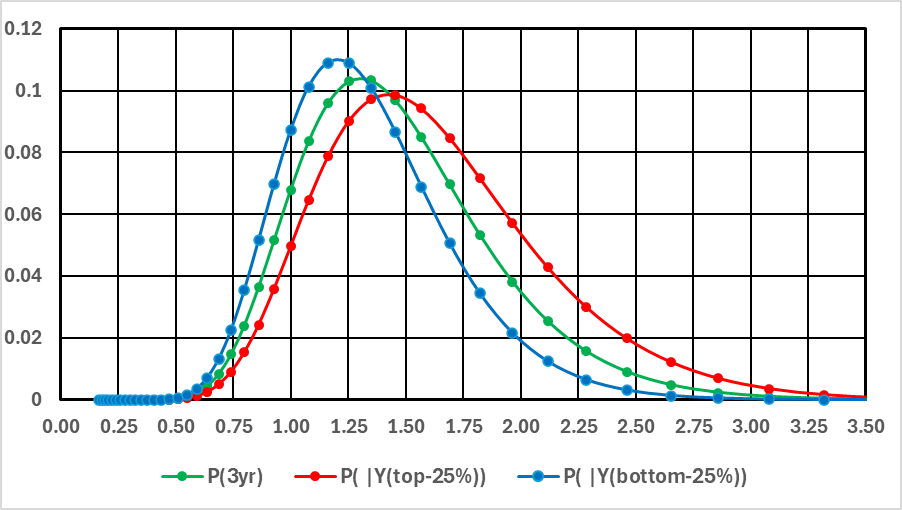}
	\caption{\footnotesize Comparison of the marginal price distribution $\mbb{P}$ 
	and the conditional price distributions $\mbb{P}(\cdot| Y^{\rm top-25\%})$
		and $\mbb{P}(\cdot |Y^{\rm bottom-25\%})$ at 3-year point for $(\ref{Terminal-F1})$. }
	\label{fig-top-bottom-1}  
  \end{minipage}

    \vspace{4mm}
    \centering
    \begin{minipage}[t]{0.49\textwidth}
        \centering
        \includegraphics[width=\linewidth]{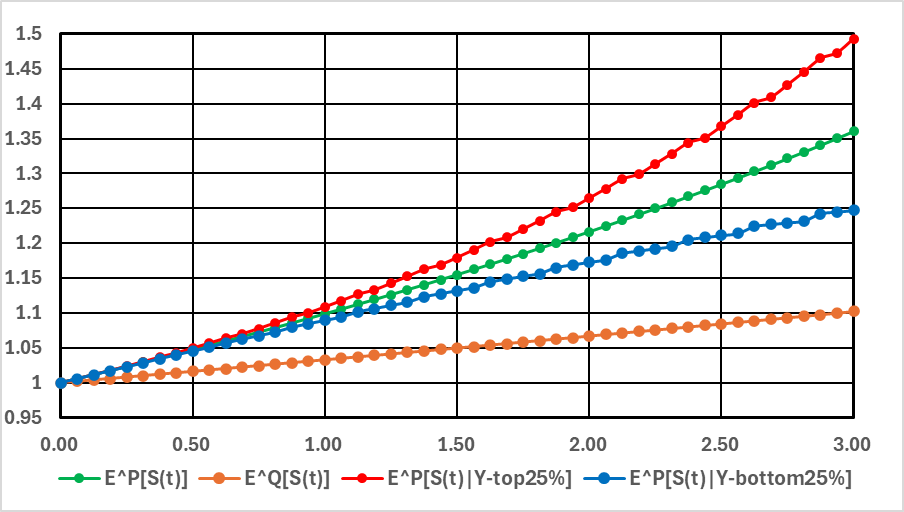}
	\caption{\footnotesize Comparison of the expected values of $S(t_n)$ under $\mbb{Q}$, $\mbb{P}$,
	$\mbb{P}(\cdot|Y^{\rm top-25\%})$, and $\mbb{P}(\cdot|Y^{\rm bottom-25\%})$ for $(\ref{Terminal-F1})$. }
	\label{fig-exp-1}
    \end{minipage}%
    \hspace{0.01\textwidth} 
    \begin{minipage}[t]{0.49\textwidth}
        \centering
        \includegraphics[width=\linewidth]{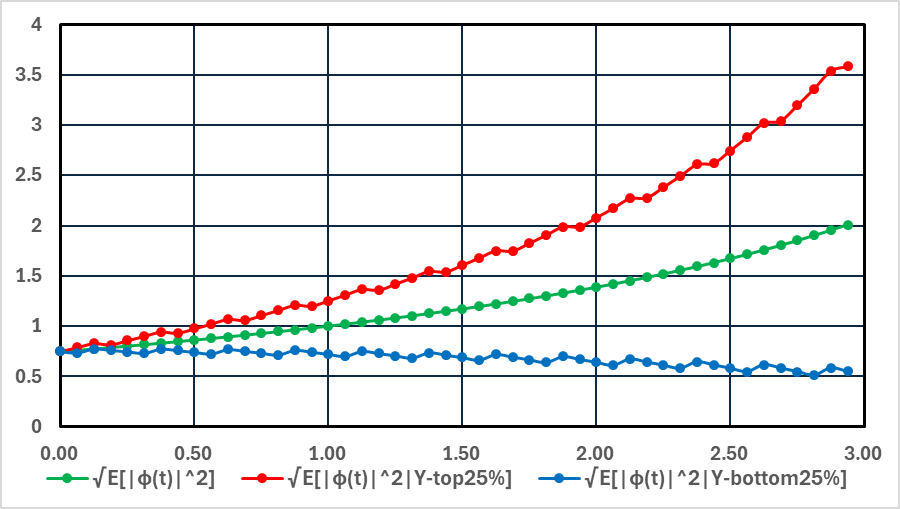}
	\caption{\footnotesize The time evolution of the expected trading volume $\mbb{E}[|\phi^{1,*}(t)|^2]^\frac{1}{2}$, 
	$\mbb{E}[|\phi^{1,*}(t)|^2|Y^{\rm top-25\%}]^\frac{1}{2}$, and $\mbb{E}[|\phi^{1,*}(t)|^2|Y^{\rm bottom-25\%}]^\frac{1}{2}$ 
	for $(\ref{Terminal-F1})$.}
	\label{fig-phi-norm-1}
    \end{minipage}
\end{figure}

Let us first assume that there is no external oder flow $L_n\equiv 0$, and that the 
terminal liability $F$ is given by 
\be
F(S_N,Y_N,Z_N^i):=C-3S_N Y_N Z_N^i, 
\label{Terminal-F1}
\ee
where $C\in \mbb{R}$ is an arbitrary real constant. Since the result is invariant to a constant shift,
one may adjust the constant $C$, if necessary,  to make the liability positive.
From the discussion in Section~\ref{sec-implications}, 
we expect that the equilibrium price distribution for this liability will yield a positive excess return.
We can verify this expectation by observing Figure~\ref{fig-marginal-1},
which presents the comparison of the marginal price distributions under the equilibrium measure $(\mbb{P})$
and the risk-neutral measure $(\mbb{Q})$ at 1-year and 3-year points.

We can also provide the conditional price distribution $\mbb{P}(S_n\in A|Y_n=y), ~\forall A\subset \cals_n$ for each  $y\in \caly_n$.  
At the three-year point $(n=48)$, the value of $Y$ marking the 75th percentile  $(Y^{\rm top-25\%})$ 
is equivalent to a total of 36 up moves, while the 25th percentile $(Y^{\rm bottom-25\%})$ is 
equivalent to 12 up moves.
 Figure~\ref{fig-top-bottom-1} compares the conditional distributions $\{\mbb{P}(S_n=s|Y^{\rm top-25\%})$, $\mbb{P}(S_n=s|Y^{\rm bottom-25\%}), ~\forall s\in \cals_n\}$ with the marginal distribution $\mbb{P}(\cdot)$ at the three-year point.
As expected from the functional form in $(\ref{Terminal-F1})$, the deviations from the risk-neutral distribution 
are positive and become larger for the larger value of  $Y$. 
If $Y$ is a stochastic process possessing large jumps (instead of the diffusion-like process in this section), 
which is the case, for example,  for regime-switching models, 
we expect a substantial sudden shift of the transition probabilities and hence the equilibrium excess return.

\begin{figure}[H]
\vspace{0mm}
    \centering
    \begin{minipage}[t]{0.49\textwidth}
        \centering
        \includegraphics[width=\linewidth]{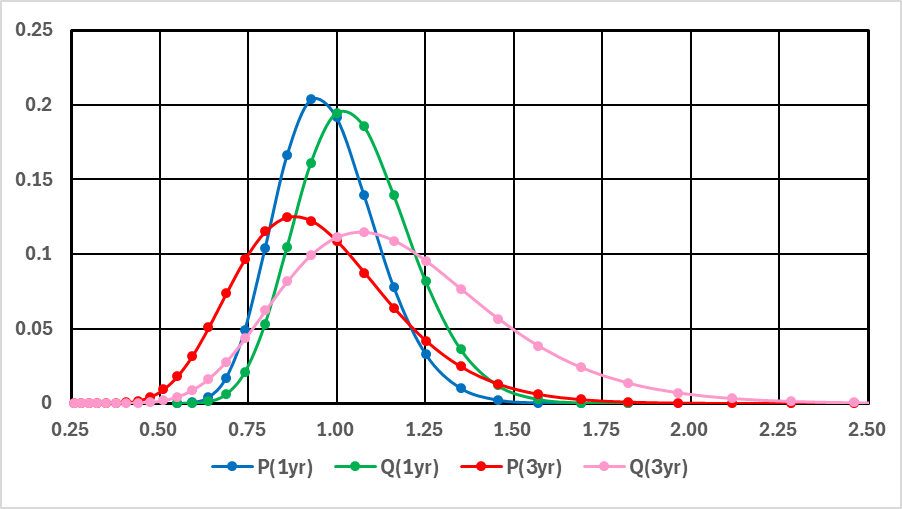}
	\caption{\footnotesize Comparison of the marginal price distributions under the equilibrium measure $(\mbb{P})$ 
	and the risk-neutral measures $(\mbb{Q})$
	at 1-year and 3-year points for $(\ref{Terminal-F2})$.}
	 \label{fig-marginal-2}
    \end{minipage}%
    \hspace{0.01\textwidth} 
    \begin{minipage}[t]{0.49\textwidth}
        \centering
        \includegraphics[width=\linewidth]{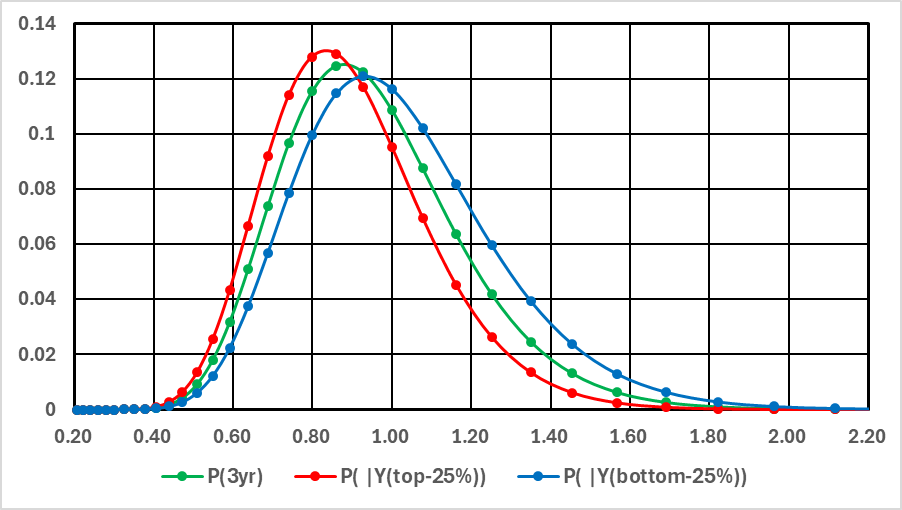}
	\caption{\footnotesize Comparison of the marginal price distribution $\mbb{P}$ 
	and the conditional price distributions $\mbb{P}(\cdot| Y^{\rm top-25\%})$
		and $\mbb{P}(\cdot |Y^{\rm bottom-25\%})$ at 3-year point for $(\ref{Terminal-F2})$. }
	\label{fig-top-bottom-2}  
  \end{minipage}

    \vspace{4mm}
    \centering
    \begin{minipage}[t]{0.49\textwidth}
        \centering
        \includegraphics[width=\linewidth]{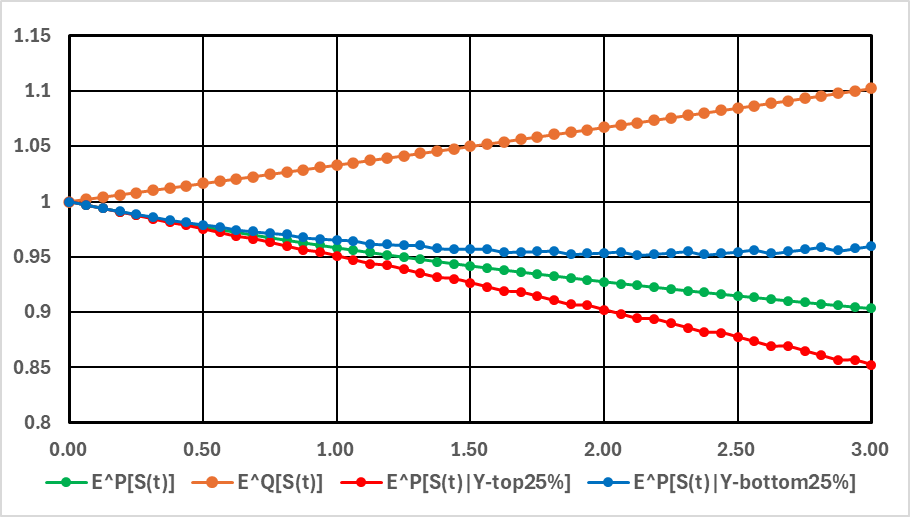}
	\caption{\footnotesize 
	Comparison of the expected values of $S(t_n)$ under $\mbb{Q}$, $\mbb{P}$,
	$\mbb{P}(\cdot|Y^{\rm top-25\%})$, and $\mbb{P}(\cdot|Y^{\rm bottom-25\%})$ for $(\ref{Terminal-F2})$. }
	\label{fig-exp-2}
    \end{minipage}%
    \hspace{0.01\textwidth} 
    \begin{minipage}[t]{0.49\textwidth}
        \centering
        \includegraphics[width=\linewidth]{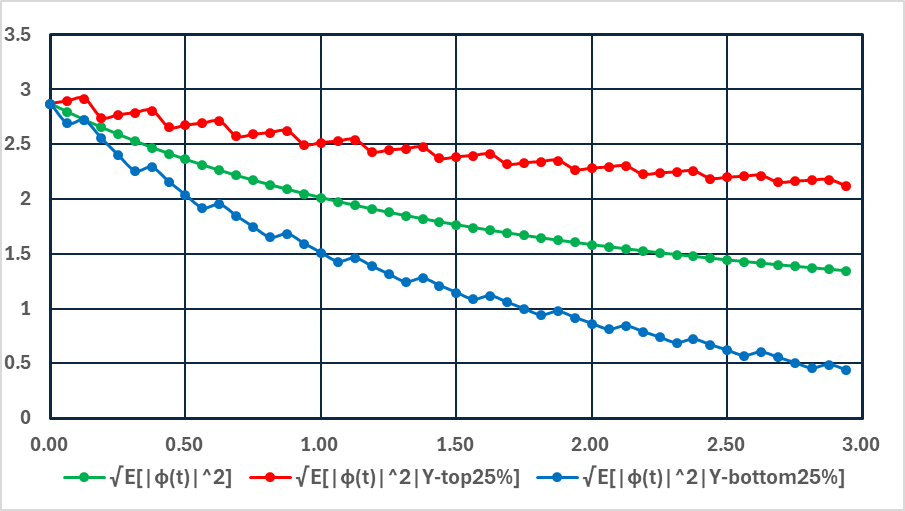}
	\caption{\footnotesize 
The time evolution of the expected trading volume $\mbb{E}[|\phi^{1,*}(t)|^2]^\frac{1}{2}$, 
	$\mbb{E}[|\phi^{1,*}(t)|^2|Y^{\rm top-25\%}]^\frac{1}{2}$, and $\mbb{E}[|\phi^{1,*}(t)|^2|Y^{\rm bottom-25\%}]^\frac{1}{2}$ 
for $(\ref{Terminal-F2})$.}
	\label{fig-phi-norm-2}
    \end{minipage}
\end{figure}
In Figure~\ref{fig-exp-1}, we provide the time-evolution of the expected value of the stock price under the equilibrium $(\mbb{P})$ 
and the risk-neutral $(\mbb{Q})$ measures $(\mbb{E}^{\mbb{P}}[S(t)], \ex^{\mbb{Q}}[S(t)])$.
We also include the conditional expectation values 
$(\mbb{E}^{\mbb{P}}[S(t)|Y^{\rm top-25\%}], \mbb{E}^{\mbb{P}}[S(t)|Y^{\rm bottom-25\%}])$
for $Y$ in the 75th $(Y^{\rm top-25\%})$ and 25th percentiles $(Y^{\rm bottom-25\%})$, choosing the nearest nodes at each time.
Since the risk-free rate is $r=3.3\%$ per annum, one can observe from Figure~\ref{fig-exp-1} that the 
excess return is roughly $8\%$ for the equilibrium distribution $\mbb{P}$,
and roughly $13\%$ and $5\%$ per annum, conditional on $Y^{\rm top-25\%}$ and $Y^{\rm bottom-25\%}$, respectively.
Figure~\ref{fig-phi-norm-1} gives the time-evolution of the expected trading volume $\mbb{E}^\mbb{P}[|\phi^{1,*}(t)|^2]^\frac{1}{2}$
under the equilibrium distribution,  as well as the conditional expectations $(\mbb{E}^\mbb{P}[|\phi^{1,*}(t)|^2|Y^{\rm top-25\%}]^\frac{1}{2},
\mbb{E}^\mbb{P}[|\phi^{1,*}(t)|^2|Y^{\rm bottom-25\%}]^\frac{1}{2})$.
The slightly irregular zigzag behavior observed in Figures~\ref{fig-exp-1} and \ref{fig-phi-norm-1}
is an artifact of our selection of the nearest nodes when determining the top and bottom 25th percentiles for $Y$ at each time step.

In Figures~\ref{fig-marginal-2}, \ref{fig-top-bottom-2},  \ref{fig-exp-2} and \ref{fig-phi-norm-2},  we present 
the corresponding results for the different liability function
\be
F(S_N,Y_N,Z_N^i):=C+3S_N Y_N Z_N^i 
\label{Terminal-F2}
\ee
which exhibits the opposite sign of sensitivity to the stock price. As expected, we observe that the excess return becomes negative in this case.

\begin{figure}[H]
\vspace{2mm}
    \centering
    \begin{minipage}[t]{0.498\textwidth}
        \centering
        \includegraphics[width=\linewidth]{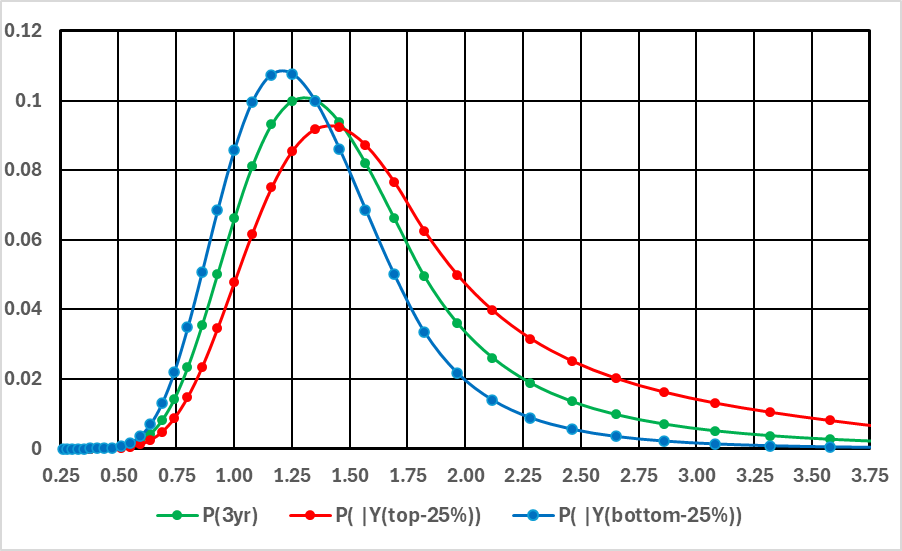}
    \end{minipage}%
    \hspace{-0.005\textwidth} 
    \begin{minipage}[t]{0.498\textwidth}
        \centering
        \includegraphics[width=\linewidth]{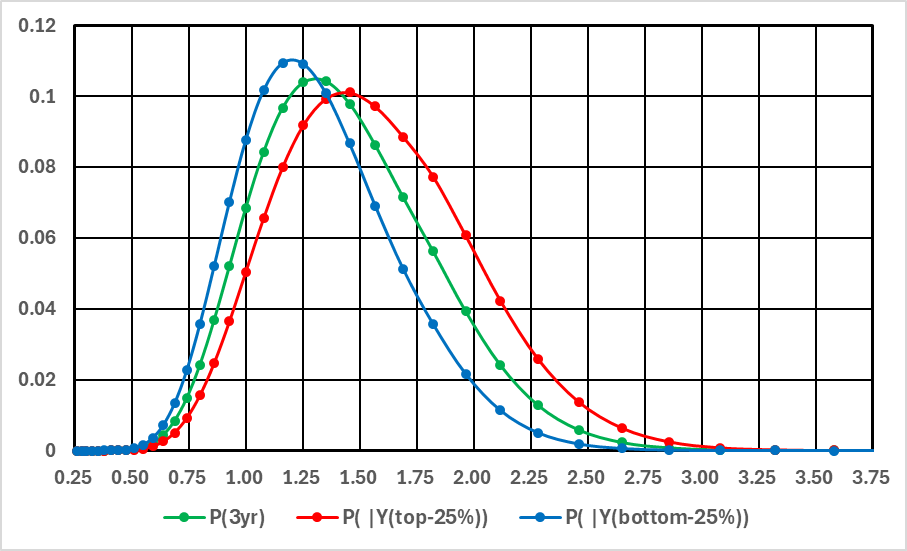}
    \end{minipage}
    \caption{\footnotesize
Comparison of the marginal price distribution $\mbb{P}$ and the conditional price distributions $\mbb{P}(\cdot | Y^{\rm top-25\%})$
and $\mbb{P}(\cdot | Y^{\rm bottom-25\%})$ at 3-year point. $F$ is given by $(\ref{Terminal-F1})$ and the 
external order flow is equal to $L_n(s)=7\max(s-1.75,0)$ (i.e., positive supply) in the left panel
and $L_n(s)=-7\max(s-1.75,0)$ (i.e., positive demand) in the right panel.}
\label{fig-T1-L}
\end{figure}

Now, we introduce the external order flow.  We continue to use the same parameter values from Table~\ref{tab-param-1}
and the liability function defined by $(\ref{Terminal-F1})$, but now including an external order flow (without $y$-dependence):
\be
L_n(s):=a \max(s-c,0).
\ee
We set $c=1.75$ and consider two scenarios, one is $a=7$  and the other is $a=-7$.  In the former case, there is positive supply of the stock 
when the stock price is very large $s>1.75$, and in the latter case there is positive demand (i.e., negative supply) when $s>1.75$.
In Figure~\ref{fig-T1-L}, we compare the marginal as well as conditional price distribution as in Figure~\ref{fig-top-bottom-1}
with the positive external order flow in the left panel and the negative one in the right panel.
We can clearly observe that the positive supply generates a heavy right tail in the equilibrium distributions
and that the positive demand causes the opposite. 
\subsection{Recursive utility}
\label{sec-num-recursive}
\begin{figure}[H]
\vspace{1mm}
    \centering
    \begin{minipage}[t]{0.48\textwidth}
        \centering
        \includegraphics[width=\linewidth]{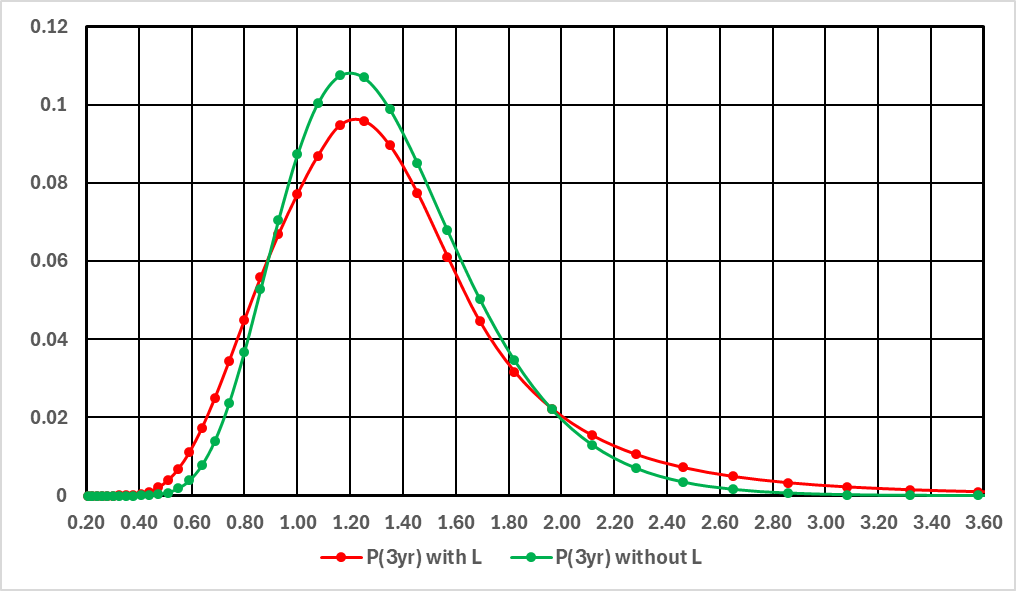}
	\caption{\footnotesize Comparison of the stock price distributions at 3-year point with $L_n$ given by $(\ref{eq-Lext})$ and $L_n\equiv 0$}
	 \label{fig-R1-LnL}
    \end{minipage}%
    \hspace{0.01\textwidth} 
    \begin{minipage}[t]{0.5\textwidth}
        \centering
        \includegraphics[width=\linewidth]{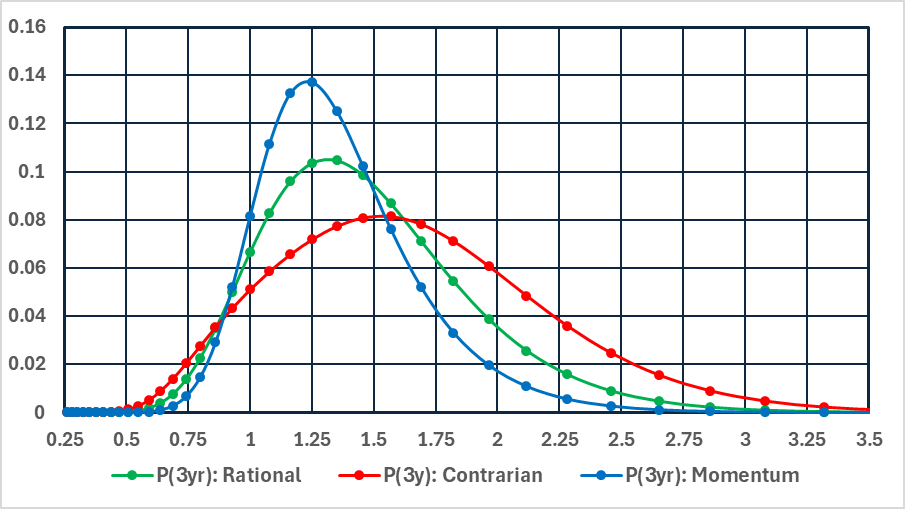}
	\caption{\footnotesize Comparison of the marginal price distribution at 3-year point with the rational agents,  those with 
contrarian-bias $(\varpi_n^c)$,  and with momentum-bias  $(\varpi_n^m)$ defined in $(\ref{eq-subjective-bias})$. }
	\label{fig-R-subjective}  
  \end{minipage}
\end{figure}

We now consider the recursive utility model discussed in Section~\ref{sec-recursive}.
The model of stochastically biased agents in Section~\ref{sec-subjective} is treated
only in Figure~\ref{fig-R-subjective}.
For numerical ease, we assume no path-dependence.
The liability $F$ is thus assumed to depend solely on the terminal stock price $S_N$,
while the incremental endowments $g_n$ depend solely on the current price $S_n$.
See Remark~\ref{remark-path}
for the corresponding analytic solutions.
We use the same models for $(S_n, Y_n,Z_n^i)_{n=0}^N$ and $\gamma_i$ as in Section~\ref{sec-num-terminal}.
$\psi_i$ is assumed to be uniformly distributed over the $(N_\psi+1)$ discrete values given by
\be
\psi_i(k_\psi):=\ul{\psi}+(\ol{\psi}-\ul{\psi})k_\psi/N_\psi, \quad k_\psi=0, \cdots, N_\psi. \nn
\ee
For simplicity, we assume that the time preference coefficient $\del_i:=\exp(-\rho \Del)$ takes a common value across the agents.
Moreover, we assume that $\zeta_i$ and $\psi_i$ are related by
$
\psi_i/\zeta_i =a_\zeta, \nn
$
where $a_\zeta$ is a positive constant common across the agents. We shall use the parameter $a_\zeta$ to control the ratio 
$
\eta_{n-1}^i/\eta_n^i \simeq\psi_i/\zeta_i. \nn
$
See the discussion in Section~\ref{sec-rec-implication} for the role of this ratio.
The parameter values to be used throughout this subsection (except the last example) are summarized in Table~\ref{tab-param-2} below:
\begin{table}[h]
\footnotesize
    \centering
    \begin{tabular}{c c c c c c c c c c c c c c c c c c c} 
        \toprule
        \text{parameter} &  $\ul{\gamma}$ &  $\ol{\gamma}$ &  $N_\gamma$ & $\ul{\psi}$ 
& $\ol{\psi}$ & $N_\psi$ & $\rho$ 
& $z_0$ & $\sigma_z$ & $p_z$ &  $Y_0$ & $\sigma_y$ & $p_y$ 
		& $S_0$ & $\sigma$  & $r$ & $T$ & $N$ \\
        \midrule
	  \text{value} & 0.4 & 1.6 & 3 & 0.5 &1.5& 2 & 5.0\% & 1.0 & 12\% & 0.5 & 1.0 & 12\% & 0.5 &1.0 & 15\%&  3.3\% & 3yr &48\\
        \bottomrule
    \end{tabular}
    \caption{parameter values}
  \label{tab-param-2}
\end{table}

\begin{figure}[H]
\vspace{-4mm}
    \centering
    \begin{minipage}[t]{0.498\textwidth}
        \centering
        \includegraphics[width=\linewidth]{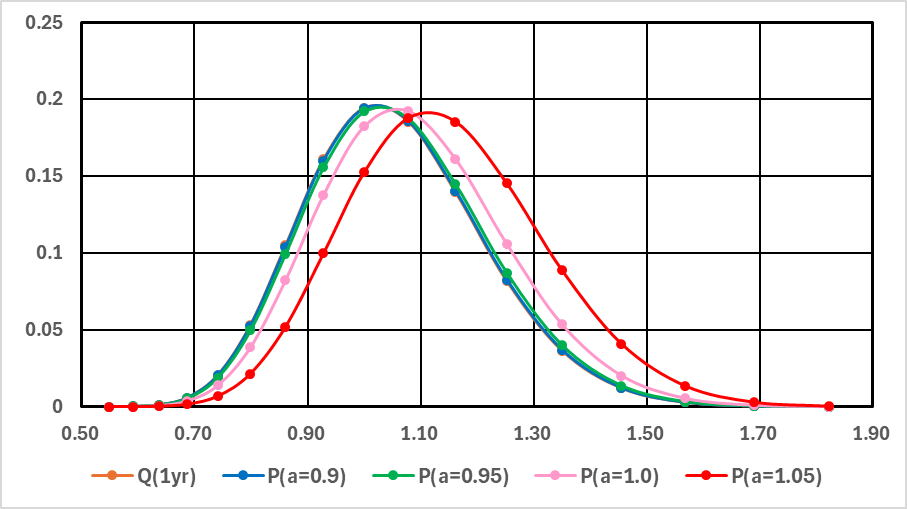}
    \end{minipage}%
    \hspace{-0.005\textwidth} 
    \begin{minipage}[t]{0.498\textwidth}
        \centering
        \includegraphics[width=\linewidth]{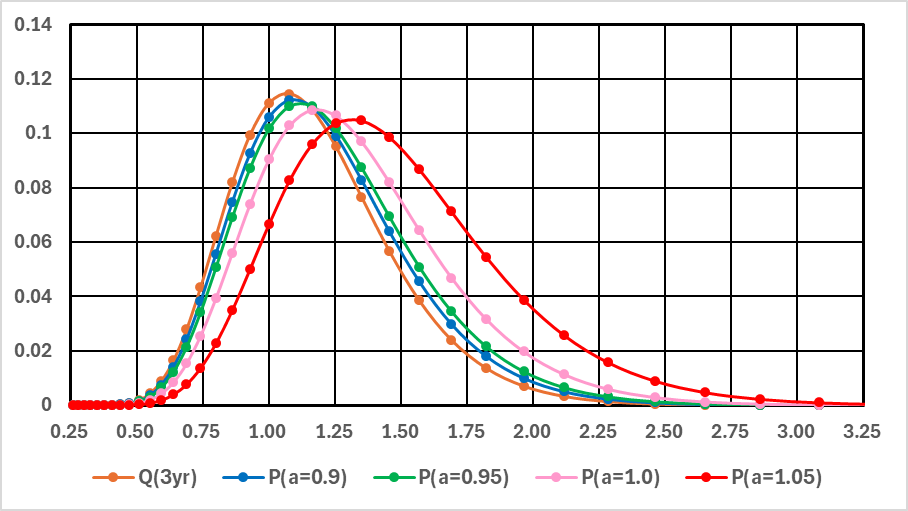}
    \end{minipage}
    \caption{\small Comparison of the risk-neutral as well as the equilibrium marginal price distributions with $a_\zeta=0.9, 0.95, 1.0, 1.05$ at 1-year (left panel) and 3-year (right panel) points.}
\label{fig-R1-1}

\vspace{2mm}
\begin{minipage}[t]{\textwidth}
\centering
    \includegraphics[width=0.5\textwidth]{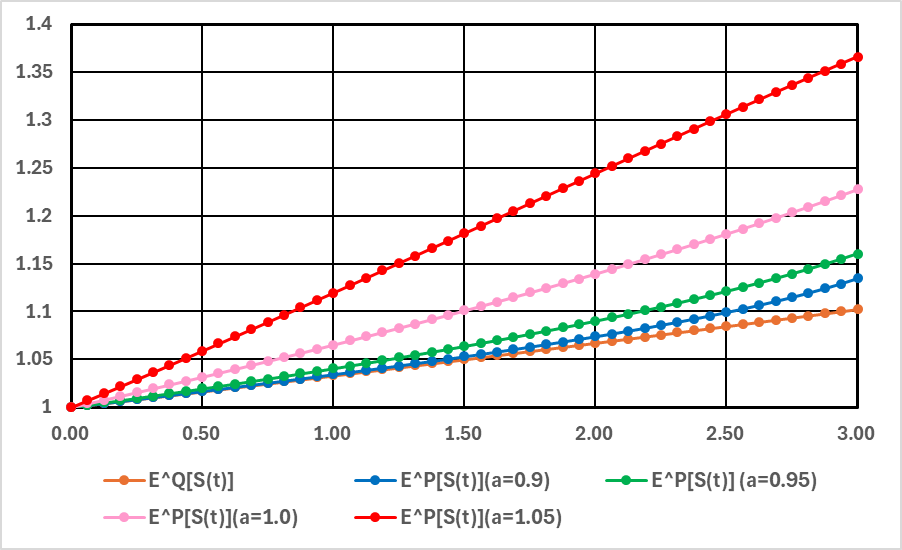}
    \caption{\small Comparison of the expected values of $S(t_n)$ under the risk-neutral as well as the equilibrium price distributions with 
$a_\zeta=0.9, 0.95, 1.0, 1.05$.}
    \label{fig-R1-variety}
\end{minipage}
\end{figure}

Since the effects of the stochastic liability and incremental endowments on the equilibrium price distributions
are as expected from the results in the previous subsection, let us first concentrate on 
the effect of the ratio $a_\zeta=\psi_i/\zeta_i$. 
We set $L_n\equiv 0, ~\forall n$ and define the liability function 
and the incremental endowments in the following way:
\bea
\label{num-functions}
&&F(S_N,Y_N,Z_N^i):=C-2S_NY_N Z_N^i, \\
&&g_n(S_n,Y_n,Z_n^i):= C^\prime+1.5 \Del S_n Y_n Z_n^i, \quad 1\leq n\leq N.
\label{num-endow}
\eea
Here, $C, C^\prime$ are arbitrary constants irrelevant for the equilibrium distributions.

In Figure~\ref{fig-R1-1}, we plot the risk-neutral as well as the equilibrium marginal price distributions
with 4 different values of the ratio $a_\zeta=0.9, 0.95, 1.0, 1.05$ at 1-year (left panel) and 3-year (right panel) points.
Figure~\ref{fig-R1-variety} provides the time evolution of the expected value of the stock price $S(t_n)$ 
for each case.  As inferred from the discussion in  Section~\ref{sec-rec-implication}, 
the deviations from the risk-neutral distribution become smaller as $a_\zeta$ decreases.
This effect (smaller deviations) is more pronounced in earlier periods.
We can observe that the value of $a_\zeta=\psi_i/\zeta_i$ can efficiently control 
the level of the excess return without changing the other parameters. 
More specifically, higher values of $a_\zeta$ put more weight on the
continuation utilities relative to  current consumptions, thus increasing the agents' hedge needs, 
and leads to higher excess returns.

Next, we set the incremental endowments $g_n\equiv 0$ while keeping the 
liability function the same as in $(\ref{num-functions})$.
We are now going to study the effects of an external order flow defined by
\be
\label{eq-Lext}
L(S_n):=8\max (S_n-1.6, 0)-8 \max(1.1-S_n,0),  \quad 0\leq n\leq N-1.
\ee
This demonstration aims to show how flexibly the shape of equilibrium distributions can change.
The definition in $(\ref{eq-Lext})$ implies a positive supply (i.e.,  sell orders from other groups) of the stock when the price $S_n$ is high
and a positive demand (i.e., buy orders from other groups) when the price is low.
In Figure~\ref{fig-R1-LnL}, with $a_\zeta=1.07$, 
we compare the equilibrium price distribution at 3-year point with $L_n$ given by $(\ref{eq-Lext})$
and that with $L_n\equiv 0$.  It shows that the existence of the external order flow $(\ref{eq-Lext})$
makes the equilibrium distribution fat-tailed in both directions, which is as expected by the analysis 
made in the last paragraph of Section~\ref{sec-implications}.
This suggests that the existence of individual investors behaving as a {\bf contrarian} 
makes the equilibrium price distribution fat-tailed, which is an interesting example to demonstrate
that microstructures of the market impact on the tail distributions.
Corresponding modifications in the terminal liability $F$ and/or the incremental endowments $g_n$ would yield similar results.

As a related analysis, we explore the impact of non-rational agents on the equilibrium price distribution, utilizing the subjective 
measure framework developed in Section~\ref{sec-subjective}. We investigate how agents' biased estimates of the transition probabilities affect
the equilibrium distribution. We define the contrarian-bias ($\varpi_n^c$) and momentum-bias ($\varpi_n^m$) as follows:
\be
\label{eq-subjective-bias}
\begin{split}
\varpi_n^c(S_n,Z_n^i)=0.8\vee \frac{S_0 \beta^n}{S_n}\frac{Z_0}{Z_n^i}\wedge 1.2, \quad 
\varpi_n^m(S_n,Z_n^i)=0.8\vee \frac{S_n}{S_0\beta^n}\frac{Z_n^i}{Z_0}\wedge 1.2, 
\end{split}
\ee
where the subjective biases are constrained  to be within $\pm 20\%$.
We use parameter values taken from Table~\ref{tab-param-2}, $a_\zeta=1.05$, and $(\ref{num-functions})$ and $(\ref{num-endow})$
for the terminal liability and the incremental endowments.
Figure~\ref{fig-R-subjective} compares the equilibrium distributions 
for the three agent types: rational, contrarian-biased, and momentum-biased agents.
As discussed in Section~\ref{sec-mfe-subjective}, the agents with contrarian-bias produce
a fat-tailed equilibrium price distribution and those with momentum-bias a thin-tailed one.

In the last numerical example, we examine the effect of $\sigma_z$, the volatility of the process $(Z^i_n)$,
on trading volume. This volume is quantified by the standard deviation of the stock position among the agents,
which is $\ex^1[|\phi^{i,*}_t|^2]^\frac{1}{2}$, as discussed in Section~\ref{sec-implications}.
To highlight the effect of $\sigma_z$, we reduce the variation in $(\gamma_i,\psi_i,\zeta_i)$.
The parameter values we use are summarized in Table~\ref{tab-param-3} below:
\begin{table}[H]
\footnotesize
    \centering
    \begin{tabular}{c c c c c c c c c c c c c c c c c c c} 
        \toprule
        \text{parameter} &  $\ul{\gamma}$ &  $\ol{\gamma}$ &  $N_\gamma$ & $\ul{\psi}$ 
& $\ol{\psi}$ & $N_\psi$ & $\rho$ 
& $z_0$ & $a_\zeta$ & $p_z$ &  $Y_0$ & $\sigma_y$ & $p_y$ 
		& $S_0$ & $\sigma$  & $r$ & $T$ & $N$ \\
        \midrule
	  \text{value} & 0.95 & 1.05 & 2 & 0.95 &1.05& 2 & 5.0\% & 1.0 & 1.02 & 0.5 & 1.0 & 12\% & 0.5 &1.0 & 15\%&  3.3\% & 3yr &48\\
        \bottomrule
    \end{tabular}
    \caption{value of parameters for Figure~\ref{fig-R1-sigmaZ}.}
  \label{tab-param-3}
\end{table}
\noindent
We put $L_n\equiv 0$ and use the liability and the incremental endowments defined by $(\ref{num-functions})$ and $(\ref{num-endow})$. 

\begin{figure}[h]
\vspace{1mm}
    \centering
    \begin{minipage}[t]{0.496\textwidth}
        \centering
        \includegraphics[width=\linewidth]{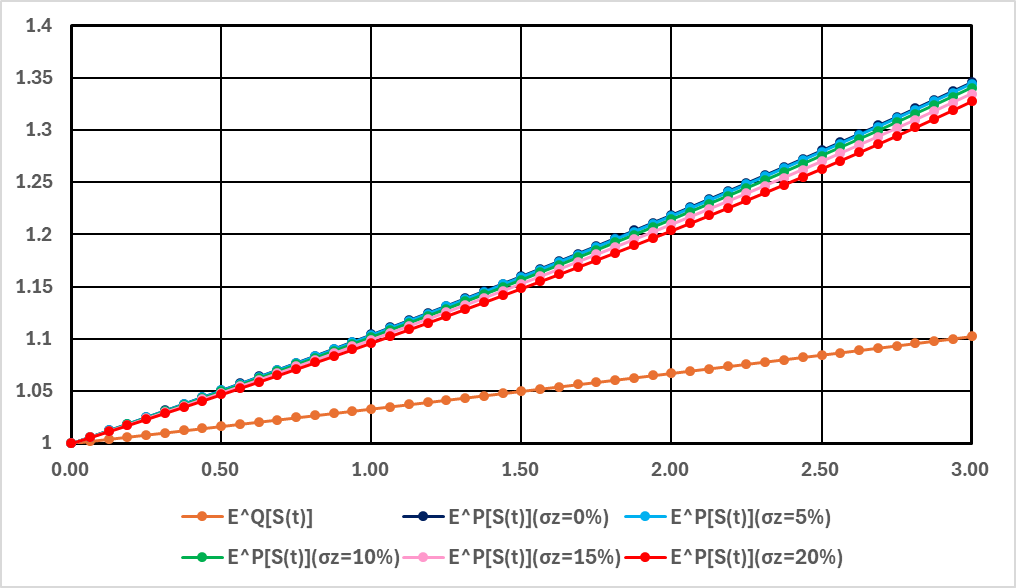}
    \end{minipage}%
    \hspace{0\textwidth} 
    \begin{minipage}[t]{0.496\textwidth}
        \centering
        \includegraphics[width=\linewidth]{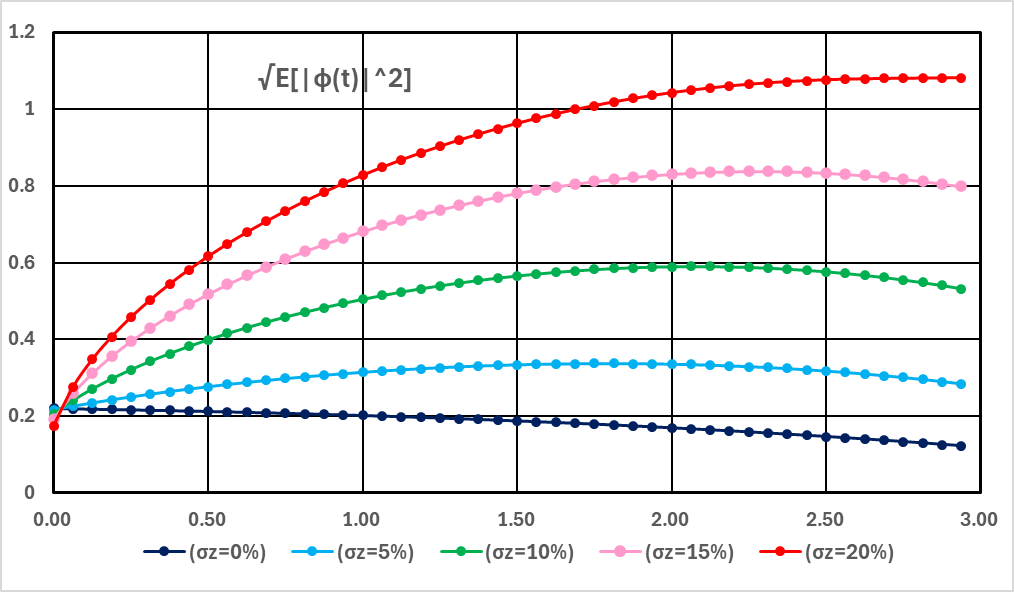}
    \end{minipage}
    \caption{\small Left panel: Comparison of the expected value of $S(t_n)$ under the risk-neutral as well as the equilibrium price distributions with 
$\sigma_z=0\%, 5\%, 10\%, 15\%, 20\%$. 
Right panel: Comparison of the trading volumes $\ex^{\mbb{P}}[|\phi^{1,*}(t)|^2]^\frac{1}{2}$ with $\sigma_z=0\%, 5\%, 10\%, 15\%, 20\%$.}
\label{fig-R1-sigmaZ}
\end{figure}
In the right panel of Figure~\ref{fig-R1-sigmaZ}, we have plotted the evolution of the trading volume $\ex^{\mbb{P}}[|\phi^{1,*}(t)|^2]^\frac{1}{2}$
for 5 different volatilities of the process $(Z^i_n)$: $\sigma_z=0\%,~5\%, ~10\%, ~15\%, ~20\%$.
We observe that the trading volume increases with the volatility $\sigma_z$.  
The non-zero trading volume,  even when $\sigma_z=0$,  stems from the non-zero variation in the risk-aversion coefficients.
The near-identical trading volume in the earliest period is a consequence of  our assumption that the agents have the common 
initial value $Z_0^i\equiv z_0=1,~\forall i\in \mbb{N}$. This  assumption is made solely for numerical convenience.

In the left panel of Figure~\ref{fig-R1-sigmaZ}, we have plotted the evolution of the expected value of $S(t)$ for each case of $\sigma_z$.
 (The result for the risk-neutral measure is also plotted for reference.) 
From this result, we see that the size of excess return is almost unaffected by the volatility $\sigma_z$.
This stems from the functional form of $(\ref{num-functions})$ and $(\ref{num-endow})$,  as well as  the fact that the expectation value of $Z^i_n$ remains nearly identical across all cases. These results suggest that we can control trading volume by changing $\sigma_z$ without significantly affecting the excess return.
Although trading volume is also significantly influenced by the variation in $\calf_0^i$-measurable random variables such as $\gamma_i, \zeta_i$,  and 
$\psi_i$,  these variables may simultaneously cause a large shift in the excess return.

\subsection{Limits of  our framework and practical implications}
As our examples demonstrated, the equilibrium price distributions and resultant excess returns 
can deviate substantially from those in the risk-neutral measure, depending on the agents' characteristics. 
This deviation can lead to extreme scenarios where the required equilibrium excess return 
(or, more generally, transition probabilities) appears unrealistic---for instance, reaching levels that seem economically implausible.
As is common in price-formation frameworks, our model does not address how 
such stock price performance can be justified by the issuer's business activities or the broader 
macroeconomic environment. Our framework solely provides the required transition probabilities (and thus excess returns) 
necessary for the agents to be compensated for their risk and to achieve a market-clearing equilibrium. 
Consequently, it does not determine whether the equilibrium is achievable in practice, 
nor does it predict the consequences of its failure.
However, our framework can still offer valuable insights for regulatory bodies regarding potential market instability. 
By explicitly identifying the necessary transition probabilities, marginal/conditional price distributions, and, in particular, 
the excess returns required to maintain equilibrium, it serves as a flexible platform for analyzing 
market stability under a variety of agent-based scenarios.

\section{Concluding remarks and future research directions}
\label{sec-conclude}

In this work, we have successfully developed a numerically tractable and readily implementable framework
for the mean-field equilibrium ($\text{MFE}$) price formation problem by combining $\text{MFG}$ theory 
with the binomial tree structure. This classical simplifying device allows us to bypass the substantial mathematical and computational challenges 
inherent in continuous-time models, such as the well-posedness and numerical evaluation of $\text{FBSDEs}$ 
of conditional McKean-Vlasov type or coupled non-linear $\text{PDEs}$.
We proved the existence of the unique $\text{MFE}$ for agents with 
exponential utilities and recursive utilities of exponential-type and derived an explicit analytic formula for the stock price transition probabilities.

Our framework successfully incorporates highly general model features, 
including stochastic terminal liabilities and incremental endowments dependent on
unhedgeable common and idiosyncratic factors, as well as external order flow. 
Moreover, the analytic tractability of our approach enabled us to achieve two significant extensions 
that would be formidable problems in the continuous-time setting: 
\begin{itemize}[noitemsep]
\item We introduced multi-population heterogeneity, allowing populations to differ fundamentally in their risk characteristics, 
liabilities, and endowments. 
\item We relaxed the standard rational expectations hypothesis by modeling agents operating under subjective probability measures. 
\end{itemize}

Our results clearly show that the equilibrium distributions can substantially change their shapes in response to these inputs. 
In particular, we found that countercyclical liabilities (or cyclical endowments) increase the excess return required by the agents, 
and that the equilibrium price distributions can become fat-tailed due to either contrarian 
external order flow or contrarian biases arising from agents' subjective measures. 
Furthermore, trading volume per capita is crucially dependent on the variation in idiosyncratic factors. 
The explicit and tractable solution offers regulatory bodies a flexible platform for market stability analysis 
under various agent-based scenarios. Empirical analysis regarding these findings would constitute an important research topic.

Our method may also be applied to other asset classes, such as 
commodities and foreign exchanges, provided they can be modeled by binomial trees. 
Specifically, frameworks like the Black-Derman-Toy model ($\text{BDT}$)~\cite{BDT} 
could be adapted to analyze a mean-field equilibrium for risk-free interest rates. 
Furthermore, various practitioner techniques, such as implied binomial trees, 
can now serve as valuable tools for investigating the $\text{MFE}$ in our framework, 
advancing beyond their initial purpose. (For a comprehensive overview of general Markov 
processes in finance, see, e.g., \cite{Rieder} and \cite{Hernandez}.)

Nevertheless, there remain several important challenges waiting for further research.
First, extensions to general multinomial trees and multi-asset frameworks constitute interesting future research directions.
Although our framework remains conceptually the same, there appear several hurdles to be overcome.
\bi[noitemsep]
	\item Although it is not difficult to put appropriate assumptions so that there exists a unique optimal solution,
		 its explicit form is generally unavailable.
	\item There are more degrees of freedom in the transition probabilities than are imposed by the market-clearing
	conditions. This remaining freedom must be fixed by imposing an appropriate dependence structure among the assets.
\ei
Due to these issues, while the second point might be beneficial for flexibility, 
computational costs would be significantly higher than the single asset case, in particular, in the presence of 
common noises.  The fact that the market-clearing condition alone does not uniquely determine the price processes 
in the presence of multiple  stocks is already well known. (See,  Karatzas \& Shreve~\cite[Chapter 4]{Karatzas}.)
This is because that one can build a equivalent set of mutual funds
from the original stocks without affecting the market-clearing condition.

Second, constructing a mean-field equilibrium for agents with 
non-exponential utilities (such as power-type utilities) in an incomplete market 
remains one of the most challenging problems. This is a common issue, mirroring the challenge in the continuous-time setting.
For non-exponential utilities, the optimal trade position $\phi^{i,*}_n$
is generally dependent on the agent's wealth level at $t_n$.
Since the wealth of each agent $X_n^i$ at $t_n$ depends on the trading strategy up to $t_n$, 
the mean-field equilibrium condition leads to a complex fixed-point problem involving the coupling of the backward optimal strategy
$\phi^{i,*}_n$ and the forward wealth process $X_n^i$. 
Although we can decouple the wealth process by deliberately constructing the model so that $\phi^{i,*}_n\equiv 0$, 
the resulting model allows no trading activity in the market and is thus clearly unrealistic.

\subsection*{Acknowledgements}
The author would like to thank M. Sekine for useful discussions related to the earlier works.

\subsubsection*{Declarations of Interest and AI use}
{\footnotesize
This research did not receive any specific grant from funding agencies in the public, commercial, or not-for-profit sectors.
There are no competing interests to declare.
The author acknowledges the use of the large language model Gemini (2.5 and 3) to refine the English clarity and style in the manuscript.
After using this tool, the author reviewed and
edited the content as needed and takes full responsibility for the content of the published article.
}
\appendix
\section{Proofs of Main Results} 
\label{sec-A}

\setcounter{equation}{0}
\setcounter{theorem}{0}
\setcounter{proposition}{0}
\setcounter{lemma}{0}
\renewcommand{\theequation}{A.\arabic{equation}}
\renewcommand{\thetheorem}{A.\arabic{theorem}}
\renewcommand{\theproposition}{A.\arabic{proposition}}
\renewcommand{\thelemma}{A.\arabic{lemma}}

\subsection{Proof of Theorem~\ref{th-t1}}
\label{A-th-t1}
\begin{proof}
We proceed by backward induction.
Let $V_N:\cals_N\times \caly_N\times \calz_N\times \Gamma \rightarrow \mbb{R}$ be 
defined by $V_N(s,y,z^i,\gamma_i)=\exp\bigl(\gamma_i F(s,y,z^i)\bigr)$. Suppose that
the problem of agent-$i$ at $t=t_{n-1}$ over the interval $[t_{n-1},t_n]$, $1\leq n\leq N$,
is given by
\be
\label{th-t1-temp1}
\sup_{\phi^i}\ex^{0,i}\Bigl[-\exp\Bigl(-\gamma_i \frac{\beta^N}{\beta^n}X_n^i\Bigr)V_n(S_n,Y_n,Z_n^i,\gamma_i)|\calf_{t_{n-1}}^{0,i}\Bigr], 
\ee
with some  measurable function $V_n:\cals_n\times \caly_n\times \calz_n\times \Gamma \rightarrow \mbb{R}$
satisfying the uniform bounds $0<c_n\leq V_n\leq C_n<\infty$ for some positive constants $c_n,C_n$.
Here, the supremum is taken over $\calf_{t_{n-1}}^{0,i}$-measurable real random variables.
Consider the optimization problem conditional on the event 
$\{\omega^{0,i}\in \Omega^{0,i} : (X^i_{n-1},S_{n-1},Y_{n-1},Z_{n-1}^i, \gamma_i)=(x^i,s,y,z^i,\gamma_i)\}$
\footnote{With a slight 
abuse of notation,  we use the same symbols for the realizations of $\calf^i_0$-measurable random variables.}.
The above problem is equivalent to
\be
\begin{split}
&\inf_{\phi^i\in \mbb{R}}\ex^{0,i}\Bigl[\exp\Bigl(-\gamma_i \frac{\beta^N}{\beta^n}(\beta x^i+\phi^i R_n)\Bigr)
V_n(s\wt{R}_n,Y_n,Z_n^i,\gamma_i)\bigr|x^i,s,y,z^i,\gamma_i\Bigr]\\
&=\exp\Bigl(-\gamma_i \frac{\beta^N}{\beta^{n-1}}x^i\Bigr)\inf_{\phi^i\in \mbb{R}}
\Bigl\{p_{n-1}(s,y)\exp\Bigl(-\gamma_i \frac{\beta^N}{\beta^n}\phi^i u\Bigr)\ex^{0,i}[V_n(s\wt{u},Y_n,Z_n^i,\gamma_i)|y,z^i,\gamma_i] \\
&\qquad\qquad  +q_{n-1}(s,y)\exp\Bigl(-\gamma_i \frac{\beta^N}{\beta^n}\phi^i d \Bigr)\ex^{0,i}[V_n(s\wt{d},Y_n,Z_n^i,\gamma_i)|y,z^i,\gamma_i]\Bigr\}, \nn
\end{split}
\ee
where we have used the property given in Assumption~\ref{assumption-T1} {\rm (vi)}.
Since $d<0<u$, the optimal position $\phi^{i,*}$ is a.s.~uniquely characterized by
\be
\begin{split}
&p_{n-1}(s,y)u\exp\Bigl({-\gamma_i \frac{\beta^N}{\beta^n}\phi^{i,*}u}\Bigr)\ex^{0,i}[V_n(s\wt{u},Y_n,Z_n^i,\gamma_i)|y,z^i,\gamma_i] \\
&\quad +q_{n-1}(s,y)d \exp\Bigl({-\gamma_i \frac{\beta^N}{\beta^n}\phi^{i,*}d}\Bigr)\ex^{0,i}[V_n(s\wt{d},Y_n,Z_n^i,\gamma_i)|y,z^i,\gamma_i] =0. \nn
\end{split}
\ee
Solving this yields the expression in $(\ref{th-t1-eq1})$. This solution is well-defined and bounded because $c_n\leq V_n\leq C_n$, 
and the condition $0<p_{n-1}(s,y),q_{n-1}(s,y)<1$ holds for every $(s,y)\in \cals_{n-1}\times\caly_{n-1}$
by assumption. Note that $\log f_{n-1}$ is bounded. 
It follows that the function $V_{n-1}$ defined by $(\ref{th-t1-eq2})$ remains a bounded function 
satisfying $0<c_{n-1}\leq V_{n-1}\leq C_{n-1}<\infty$ for some positive constants
uniformly on the domain $\cals_{n-1}\times \caly_{n-1}\times \calz_{n-1}\times \Gamma$.
The value function at the previous time step $t_{n-1}$ takes the form
\be
-\exp\Bigl(-\gamma_i \frac{\beta^N}{\beta^{n-1}}X_{n-1}^i\Bigr) V_{n-1}(S_{n-1},Y_{n-1},Z_{n-1}^i,\gamma_i) \nn
\ee 
and we have recovered the problem of the same form as in $(\ref{th-t1-temp1})$.
Iterating this procedure backward from $n=N$ to $1$ yields the desired conclusion.
\end{proof}
\subsection{Proof of Theorem~\ref{th-t2}}
\label{A-th-t2}
\begin{proof}
The first claim $(\ref{th-t2-eq1})$ is a direct consequence of $(\ref{th-t1-eq1})$. The form of $p_{n-1}(s,y)$ is uniquely determined by 
solving the condition $(\ref{def-mfe-t})$.   
By substituting the resultant expression for $p_{n-1}(s,y)$ (and $q_{n-1}(s,y)$) into $(\ref{th-t1-eq1})$, we obtain $(\ref{th-t2-eq3})$.

We only need to verify that the family of transition probabilities $(p_{n-1}(s,y))_{n=1}^N$ defined in $(\ref{th-t2-eq1})$ 
satisfies the bound given by  (vi) in Assumption~\ref{assumption-T1}
while updating the functions $(f_{n-1})$ by the backward induction given in Theorem~\ref{th-t1}.
At $t=t_{N-1}$, $f_{N-1}$ is a measurable function satisfying $0<c_N\leq f_{N-1}\leq C_N<\infty $
uniformly on its domain for some positive constants $(c_N, C_N)$ by the corresponding assumption on $F$. 
Since $d<0<u$, it is easy to see $0<p_{N-1}(s,y), q_{N-1}(s,y)<1$ for every $(s,y)\in \cals_{N-1}\times \caly_{N-1}$, 
which is consistent with the condition. This implies that $\phi^{i,*}_{N-1}$ is a bounded function on 
$\cals_{N-1}\times \caly_{N-1}\times \calz_{N-1}\times \Gamma$. 
This then implies that $V_{N-1}$ given by $(\ref{th-t1-eq2})$ satisfies $0<c_{N-1}\leq V_{N-1}\leq C_{N-1}<\infty$ 
uniformly on its domain with some positive constants $(c_{N-1},C_{N-1})$.
This in turn ensures that  $f_{N-2}$ satisfies the desired properties, 
and so do $(p_{N-2}(s,y), q_{N-2}(s,y))$, $(s,y)\in \cals_{N-2}\times \caly_{N-2}$. 
Proceeding in this way, by backward induction, we get the desired consistency for every time step.

For the second claim, it suffices to show that there is some constant $\calc_{n-1}$ such that an inequality
\be
\ex \Bigl|\frac{1}{N_p}\sum_{i=1}^{N_p}\phi^{i,*}_{n-1}(s,y, Z_{n-1}^i,\gamma_i)-L_{n-1}(s,y)\Bigr|^2 \leq \frac{\calc_{n-1}}{N_p} \nn
\ee
holds for every $(s,y)\in \cals_{n-1}\times \caly_{n-1}$,  $1\leq n\leq N$. 
Using the i.i.d. property of $(Z_{n-1}^i,\gamma_i)$ and the boundedness of $f_{n-1}, L_{n-1}$ and $1/\gamma_i$, 
we can evaluate the expression $(\ref{th-t2-eq3})$ to show that there exists some constant $\calc_{n-1}$ such that
\be
\begin{split}
&\ex \Bigl|\frac{1}{N_p}\sum_{i=1}^{N_p}\phi^{i,*}_{n-1}(s,y, Z_{n-1}^i,\gamma_i)-L_{n-1}(s,y)\Bigr|^2  \\
&\leq \frac{\calc_{n-1}}{N_p^2} \ex \Bigl[\Bigl|\sum_{i=1}^{N_p}\Bigl(\frac{1}{\gamma_i}-\ex^1\Bigl[\frac{1}{\gamma_1}\Bigr]\Bigr)\Bigr|^2
+\Bigl|\sum_{i=1}^{N_p}\Bigl(\frac{\log f_{n-1}(s,y,Z_{n-1}^i,\gamma_i)}{\gamma_i}-\ex^1\Bigl[\frac{\log f_{n-1}(s,y,Z_{n-1}^1,\gamma_1)}{\gamma_1}\Bigr]
\Bigr)\Bigr|^2\Bigr]\\
&\leq \frac{\calc_{n-1}}{N_p}\ex^1\left[
\Bigl|\frac{1}{\gamma_1}-\ex^1\Bigl(\frac{1}{\gamma_1}\Bigr)\Bigr|^2+
\Bigl|\frac{\log f_{n-1}(s,y,Z_{n-1}^1,\gamma_1)}{\gamma_1}-\ex^1\Bigl[\frac{\log f_{n-1}(s,y,Z_{n-1}^1,\gamma_1)}{\gamma_1}\Bigr]\Bigr|^2\right] \nn
\end{split}
\ee
which establishes the desired result. Note that the cross terms  appearing in the second line vanish due to 
the independence of $(Z^i,\gamma_i)_{i\in \mbb{N}}$. The variances on the right-hand side of the third line 
are finite uniformly in $(s,y)\in \cals_{n-1}\times \caly_{n-1}$, owing to the boundedness of functions $f_{n-1}$ and $1/\gamma_i$.
\end{proof}
\subsection{Proof of Theorem~\ref{th-R1}}
\label{A-th-R1}
\begin{proof}
We first hypothesize that the utility $U_n^i$ at $t=t_n$ is given by the following form:
\be
U_n^i(X_n^i, \bS^n, Y_n, Z_n^i,\vr_i)=\eta_n^i X_n^i-V_n(\bS^n, Y_n, Z_n^i, \vr_i), \nn
\ee
where $V_n:\cals^n\times \caly_n\times \calz_n\times\Gamma\rightarrow \mbb{R}$
is a bounded measurable function and $\eta_n^i$ is an $\calf_0^i$-measurable strictly positive and bounded random variable.  
The hypothesis trivially holds at the terminal point
with $V_N(\bS^N,Y_N,Z_N^i,\vr_i)=F(\bS^N,Y_N,Z_N^i)$ and $\eta_N^i\equiv 1$.
We shall show by induction that our hypothesis holds for every period.
Under the hypothesis, the problem for agent-$i$ at $t_{n-1}$ over the interval $[t_{n-1},t_n]$ 
is to find an $\calf_{t_{n-1}}^{0,i}$-measurable strategy $(\phi^i,c^i)$
that solves
\be
\label{problem-R-intermediate}
\inf_{(\phi^i,c^i)}\Bigl\{ \exp({-\zeta_i c^i})\Del+\del_i \exp\Bigl(\frac{\psi_i}{\gamma_i}\log\Bigl(\ex^{0,i}
\bigl[\exp({-\gamma_i U_n^i (X_n^i,\bS^n,Y_n,Z_n^i,\vr_i)})|\calf_{t_{n-1}}^{0,i}\bigr]\Bigr)\Bigr)\Bigr\}. 
\ee

\bigskip
We consider the problem on the set $\{\omega^{0,i}\in \Omega^{0,i} : (X_{n-1}^i,\bS^{n-1},Y_{n-1},Z_{n-1}^i, \vr_i)=(x^i,\bs,y,z^i,\vr_i)\}$.
By Assumption~\ref{assumption-R1} (vi), $(\ref{Y-conditional-2})$, and the above hypothesis, we have
\be
\begin{split}
&\ex^{0,i}\Bigl[\exp\bigl({-\gamma_i U_n^i(X_n^i,\bS^n,Y_n,Z_n^i,\vr_i)}\bigr)|x^i,\bs,y^i,z^i,\vr_i\Bigr]\\
&=\ex^{0,i}\Bigl[\exp\Bigl(-\gamma_i \eta_n^i \bigl(\beta (x^i-c^i\Del)+\phi^i R_n+g_n(\bS^n,Y_n,Z_n^i)\bigr)+\gamma_i V_n(\bS^n,Y_n,Z_n^i,\vr_i)
\Bigr)|\bs,y,z^i,\vr_i\Bigr] \\
&=e^{-\gamma_i \eta_n^i \beta (x^i-c^i\Del)}\Bigl\{p_{n-1}(\bs,y)e^{-\gamma_i \eta_n^i\phi^i u}
\ex^{0,i}\bigl[\exp({\gamma_i[V_n((\bs \wt{u})^n,Y_n,Z_n^i,\vr_i)-\eta_n^i g_n ((\bs \wt{u})^n, Y_n,Z_n^i)]})|y,z^i,\vr_i\bigr]\\
&\qquad\qquad +q_{n-1}(\bs,y)e^{-\gamma_i \eta_n^i \phi^i d}\ex^{0,i}\bigl[\exp({\gamma_i[V_n((\bs \wt{d})^n,Y_n,Z_n^i,\vr_i)
-\eta_n^i g_n((\bs \wt{d})^n,Y_n,Z_n^i)]})|y,z^i,\vr_i\bigr]\Bigr\}. \nn
\end{split}
\ee
Thus the problem $(\ref{problem-R-intermediate})$ can be restated as
\be
\begin{split}
&\inf_{(\phi^i,c^i)}\Bigl\{\exp({-\zeta_i c^i})\Del+\del_i \exp({-\psi_i \eta_n^i \beta (x^i-c^i\Del)})\\
&\times \exp\Bigl(\frac{\psi_i}{\gamma_i}\log\Bigl\{ p_{n-1}(\bs,y)e^{-\gamma_i\eta_n^i\phi^i u}\ex^{0,i}
\bigl[\exp({\gamma_i[V_n((\bs \wt{u})^n,Y_n,Z_n^i,\vr_i)-\eta_n^i g_n((\bs \wt{u})^n,Y_n,Z_n^i)]})|y,z^i,\vr_i\bigr] \\
&\qquad\qquad+q_{n-1}(\bs,y)e^{-\gamma_i \eta_n^i \phi^i d}\ex^{0,i}\bigl[\exp({\gamma_i[V_n((\bs \wt{d})^n,Y_n,Z_n^i,\vr_i)-\eta_n^i g_n((\bs \wt{d})^n, Y_n,Z_n^i)]})|y,z^i,\vr_i\bigr]\Bigr\} \Bigr)\Bigr\}. \nn
\end{split}
\ee

The optimization over $(\phi^i,c^i)$ can now be solved separately. Since $d<0<u$, the optimal $\phi^{i,*}$ 
is characterized uniquely by
\be
\begin{split}
0&=p_{n-1}(\bs,y) u e^{-\gamma_i\eta_n^i\phi^i u}  \ex^{0,i}
\bigl[\exp({\gamma_i[V_n((\bs \wt{u})^n,Y_n,Z_n^i,\vr_i)-\eta_n^i g_n((\bs \wt{u})^n,Y_n,Z_n^i)]})|y,z^i,\vr_i\bigr] \\
&+q_{n-1}(\bs,y)d e^{-\gamma_i \eta_n^i \phi^i d}\ex^{0,i}\bigl[\exp({\gamma_i[V_n((\bs \wt{d})^n,Y_n,Z_n^i,\vr_i)
-\eta_n^i g_n((\bs \wt{d})^n, Y_n,Z_n^i)]})|y,z^i,\vr_i\bigr], \nn
\end{split}
\ee
which gives the desired solution $(\ref{th-R1-phinm1})$ for $\phi^{i,*}_{n-1}$ with $f_{n-1}$ defined as in $(\ref{th-R1-fnm1})$.
Due to the boundedness of $g_n$ and our hypothesis on $V_n$, 
$f_{n-1}$ satisfies the uniform bounds $0<c_n\leq f_{n-1}\leq C_n<\infty$
on its domain with some positive constants $c_n$ and $C_n$. Combined with the assumption on $(p_{n-1},q_{n-1})$
and our hypothesis on $\eta_n^i$, $\phi^{i,*}_{n-1}$ is also bounded  on its domain.

With $\wt{V}_{n-1}$ defined as in $(\ref{th-R1-Vtildenm1})$, the optimization with respect to $c^i$ reduces to 
\be
\inf_{c^i}\Bigl\{ \exp({-\zeta_i c^i})\Del+\del_i \exp\bigl({-\psi_i \eta_n^i \beta (x^i-c^i \Del)}\bigr)\exp\Bigl(\frac{\psi_i}{\gamma_i}\log \wt{V}_{n-1}
(\bs, y,z^i,\vr_i)\Bigr)\Bigr\}.\nn
\ee
By the boundedness of $V_n, g_n$ and $\phi^{i,*}_{n-1}$, the function $\wt{V}_{n-1}$ also
satisfies the uniform bounds $0<c_n\leq \wt{V}_{n-1}\leq C_n<\infty$ (with modifications on $(c_n, C_n)$ if necessary)
on its domain. Thus the optimal solution is characterized by the equation:
\be
\label{th-R1-c_dummy}
0=-\zeta_i \exp({-\zeta_i c^i})+\del_i \psi_i \eta_n^i\beta 
\exp\bigl({-\psi_i\eta_n^i\beta (x^i-c^i \Del)}\bigr)\exp\Bigl(\frac{\psi_i}{\gamma_i}\log \wt{V}_{n-1}
(\bs, y,z^i,\vr_i)\Bigr),
\ee
which gives the unique solution $c_{n-1}^{i,*}$ in $(\ref{th-R1-cnm1})$ as desired. Note that it is bounded if the wealth $x^i$
at $t_{n-1}$ is bounded. Therefore,  once our induction is complete, the spending process is shown to be bounded since $\xi_i$ takes 
values in a bounded interval $[\ul{\xi},\ol{\xi}]$.

In order to complete the induction argument,  we need to obtain the utility $U_{n-1}^i$ for the previous time step.
By $(\ref{def-RU})$, it satisfies
\be
\exp({-\zeta_i U_{n-1}^i})=\exp({-\zeta_i c^{i,*}_{n-1}})\Del+\del_i \exp\bigl({-\psi_i \eta_n^i \beta (x^i-c^{i,*}_{n-1} \Del)}\bigr)\exp\Bigl(\frac{\psi_i}{\gamma_i}\log \wt{V}_{n-1}
(\bs, y,z^i,\vr_i)\Bigr). \nn
\ee
The right-hand side of the above equality can be evaluated by using $(\ref{th-R1-c_dummy})$ as
\be
\begin{split}
&\exp({-\zeta_i c^{i,*}_{n-1}})\Del+\frac{1}{\psi_i \eta_n^i \beta}\zeta_i \exp({-\zeta_i c_{n-1}^{i,*}})=\frac{\zeta_i+\Del \psi_i\eta_n^i\beta}{\psi_i \eta_n^i\beta}
\exp({-\zeta_i c_{n-1}^{i,*}}) \\
&=\frac{\zeta_i+\Del \psi_i \eta_n^i\beta}{\psi_i \eta_n^i\beta} \exp\Bigl\{-\frac{\zeta_i  \psi_i \eta_n^i\beta}{\zeta_i+\Del \psi_i \eta_n^i\beta} x^i
+\frac{\zeta_i}{\zeta_i+\Del \psi_i \eta_n^i\beta}\log\Bigl[\frac{\del_i\psi_i\eta_n^i\beta}{\zeta_i}\exp
\Bigl(\frac{\psi_i}{\gamma_i}\log\wt{V}_{n-1}(\bs,y,z^i,\vr_i)\Bigr)\Bigr]\Bigr\}. \nn
\end{split}
\ee
It follows that the utility $U_{n-1}^i$ 
is given by 
\be
\begin{split}
U_{n-1}^i(x^i,\bs,y,z^i,\vr_i)&=\frac{\psi_i \eta_n^i \beta}{\zeta_i+\Del \psi_i \eta_n^i\beta}x^i-\frac{1}{\zeta_i+\Del \psi_i\eta_n^i\beta}
\log\Bigl\{ \frac{\del_i \psi_i \eta_n^i\beta}{\zeta_i}\exp\Bigl(\frac{\psi_i}{\gamma_i}\log\wt{V}_{n-1}(\bs,y,z^i,\vr_i)\Bigr)\Bigr\}\\
&-\frac{1}{\zeta_i}\log\Bigl(\frac{\zeta_i+\Del \psi_i\eta_n^i\beta}{\psi_i \eta_n^i\beta}\Bigr). \nn
\end{split}
\ee
on the set $\{\omega^{0,i}\in \Omega^{0,i}: (X_{n-1}^i,\bS^{n-1},Y_{n-1},Z_{n-1}^i,\vr_i)=(x^i,\bs,y,z^i,\vr_i)\}$. 
By setting the right-hand side equal to $\eta_{n-1}^i x^i-V_{n-1}(\bs,y,z^i,\vr_i)$, we obtain the desired recursive relation
for $\eta_n^i$ and $V_n$. It is now clear that $(\eta_n^i)_{n=1}^N$ are $\calf_0^i$-measurable, strictly positive and bounded,
and that $V_n$ is a bounded function on $\cals^n\times\caly_n\times \calz_n\times \Gamma\rightarrow \mbb{R}$
for every $0\leq n\leq N$.
\end{proof}

\subsection{Proof of Theorem~\ref{th-R2}}
\label{A-th-R2}
\begin{proof}
The expression for the transition probabilities $p_{n-1}(\bs,y)$ is a direct consequence of $(\ref{th-R1-phinm1})$ and $(\ref{rec-clearing})$.
By substituting this expression for $p_{n-1}(\bs,y)$ (and $q_{n-1}(\bs,y)$) into $(\ref{th-R1-phinm1})$,
we obtain $(\ref{th-R2-phinm1})$.
The rest of the proof is analogous to that of Theorem~\ref{th-t2}.
Assumptions~\ref{assumption-R2} and \ref{assumption-R3} guarantee that 
 $f_{N-1}$ satisfies the uniform bounds $c_N\leq f_{N-1}\leq C_N$ for some positive constants $c_N$ and $C_N$ 
and hence the condition (vi) in Assumption~\ref{assumption-R1} is satisfied for $(p_{N-1}, q_{N-1})$.
This implies that  $\phi^{i,*}_{N-1}$ is bounded, and consequently $\wt{V}_{N-1}$ satisfies the bounds $c_N\leq \wt{V}_{N-1}\leq C_N$
(possibly with modified constants $c_N, C_N$).  
It then follows that $f_{N-2}$ satisfies the uniform bounds $c_{N-1}\leq f_{N-2}\leq C_{N-1}$, 
and $V_{N-1}$ is bounded ($|V_{N-1}|\leq C_{N-1}$) with some positive constants $c_{N-1}$ and $C_{N-1}$.
This shows that $(p_{N-2}, q_{N-2})$ satisfies the condition (vi). 
In this way, a simple induction shows that the transition probabilities satisfy Assumption~\ref{assumption-R1} (vi)
for every period. The second claim can be proved from $(\ref{th-R2-phinm1})$
and the fact that all the relevant components are bounded.
\end{proof}

\subsection{Proof of Theorem~\ref{th-multi}}
\label{A-th-multi}
\begin{proof}
By the i.i.d.~property within each population and the decomposition $(\ref{mp-decomp})$, the condition for the mean-field equilibrium given by Definition~\ref{def-multi}
can be rewritten as
\be
\label{mfe-mp}
\sum_{p=1}^m w_p \ex^{1,p}\bigl[\phi^{1,p,*}_{n-1}(\bs,y,Z_{n-1}^{1,p},\vr_1^p)\bigr]=L_{n-1}(\bs,y) 
\ee
for every $(\bs, y)\in \cals^{n-1}\times\caly_{n-1}$, $1\leq n\leq N$. Using $(\ref{phi-m-optimal})$,
the proof can be done exactly in the same way as in Theorems~\ref{th-t2} and \ref{th-R2}.

For the second claim, using the equilibrium condition $(\ref{mfe-mp})$, it is useful to observe that
\be
\begin{split}
&\frac{1}{\caln}\sum_{p=1}^m \sum_{i=1}^{N_p}\phi^{i,p,*}_{n-1}(\bs,y,Z^{i,p}_{n-1},\vr_i^p)-L_{n-1}(\bs,y) \\
&= \sum_{p=1}^m w_p \left( \frac{1}{N_p}\sum_{i=1}^{N_p} \phi^{i,p,*}_{n-1}(\bs,y,Z^{i,p}_{n-1},\vr_i^p) - \ex^{1,p}\bigl[\phi^{1,p,*}_{n-1}(\bs,y,Z_{n-1}^{1,p},\vr_1^p)\bigr] \right). \nn
\end{split}
\ee
Since the term inside the parenthesis for each $p$ is the average of centered i.i.d.~random variables with finite variance (guaranteed by the boundedness of the effective variables), the desired convergence rate follows from the standard law of large numbers arguments used in Theorem~\ref{th-t2}.  Equivalently, one can also use the expression 
$(\ref{multi-mfe-optimal})$ with the relations $(\ref{effective-conv-1})$ and $(\ref{effective-conv-2})$. 
\end{proof}

\subsection{Proof of Theorem~\ref{th-R1P}}
\label{A-th-R1P}

\begin{proof}
The proof of this theorem follows exactly the same arguments as that of Theorem~\ref{th-R1},
with the sole modification that the objective transition probabilities $(p_{n-1}(\bs,y),q_{n-1}(\bs,y))$
are replaced by the subjective probabilities $(\mdp_{n-1}(\bs,y,z^i,\vr_i),\mdq_{n-1}(\bs,y,z^i,\vr_i))$.
Note, in particular,  that on the set $\{\omega^{0,i}\in \Omega^{0,i}: (X_{n-1}^i,\bS^{n-1},Y_{n-1},Z_{n-1}^i,\vr_i)=(x^i,\bs,y,z^i,\vr_i)\}$,
we have the following equality:
\be
\begin{split}
&\ex_\calp^{0,i}\Bigl[\exp\bigl({-\gamma_i U_n^i(X_n^i,\bS^n,Y_n,Z_n^i,\vr_i)}\bigr)|x^i,\bs,y^i,z^i,\vr_i\Bigr]\\
&=\ex^{0,i}_\calp\Bigl[\exp\Bigl(-\gamma_i \eta_n^i \bigl(\beta (x^i-c^i\Del)+\phi^i R_n+g_n(\bS^n,Y_n,Z_n^i)\bigr)+\gamma_i V_n(\bS^n,Y_n,Z_n^i,\vr_i)
\Bigr)|\bs,y,z^i,\vr_i\Bigr] \\
&=\exp({-\gamma_i \eta_n^i \beta (x^i-c^i\Del)})\\
&\quad\times \Bigl\{\mdp_{n-1}(\bs,y,z^i,\vr_i)e^{-\gamma_i \eta_n^i\phi^i u}
\ex^{0,i}\bigl[\exp({\gamma_i[V_n((\bs \wt{u})^n,Y_n,Z_n^i,\vr_i)-\eta_n^i g_n ((\bs \wt{u})^n, Y_n,Z_n^i)]})|y,z^i,\vr_i\bigr]\\
&\qquad +\mdq_{n-1}(\bs,y,z^i,\vr_i)e^{-\gamma_i \eta_n^i \phi^i d}\ex^{0,i}\bigl[\exp({\gamma_i[V_n((\bs \wt{d})^n,Y_n,Z_n^i,\vr_i)
-\eta_n^i g_n((\bs \wt{d})^n,Y_n,Z_n^i)]})|y,z^i,\vr_i\bigr]\Bigr\}. \nn
\end{split}
\ee
Here, we have used Assumption~\ref{assumption-R-P} and the coincidence of $\ex^{0,i}$ and $\ex^{0,i}_\calp$ on $\sigma(\{Y,Z^i,\vr_i\})$.
Due to the strict positivity and boundedness of $\varpi_n$, the remaining  arguments are identical to those in the proof of Theorem~\ref{th-R1}.
\end{proof}

\subsection{Proof of Theorem~\ref{th-R2-P}}
\label{A-th-R2-P}
\begin{proof}
The proof of this theorem follows exactly the same arguments as that of Theorem~\ref{th-R2}.
The expression of the objective transition probabilities can be derived as before using $(f^\pi_{n-1})$ instead of $(f_{n-1})$. 
Since $(\varpi_n)$ is strictly positive and bounded, every argument for Theorem~\ref{th-R2} is still valid.  
\end{proof}


\end{document}

%% file: Fmacro-2015.tex

\newtheorem{definition}{Definition}[section]
\newtheorem{assumption}{Assumption}[section]
\newtheorem{condition}{$[$ C}
\newtheorem{lemma}{Lemma}[section]
\newtheorem{proposition}{Proposition}[section]
\newtheorem{theorem}{Theorem}[section]
\newtheorem{remark}{Remark}[section]
\newtheorem{example}{Example}[section]
\newtheorem{corollary}{Corollary}[section]
%
\def\cala{{\cal A}}
\def\calb{{\cal B}}
\def\calc{{\cal C}}
\def\cald{{\cal D}}
\def\cale{{\cal E}}
\def\calf{{\cal F}}
\def\calg{{\cal G}}
\def\calh{{\cal H}}
\def\cali{{\cal I}}
\def\calj{{\cal J}}
\def\calk{{\cal K}}
\def\call{{\cal L}}
\def\calm{{\cal M}}
\def\caln{{\cal N}}
\def\calo{{\cal O}}
\def\calp{{\cal P}}
\def\calq{{\cal Q}}
\def\calr{{\cal R}}
\def\cals{{\cal S}}
\def\calt{{\cal T}}
\def\calu{{\cal U}}
\def\calv{{\cal V}}
\def\calw{{\cal W}}
\def\calx{{\cal X}}
\def\caly{{\cal Y}}
\def\calz{{\cal Z}}
%
\def\sskip{\hspace{0.5cm}}
\def\simleq{ \raisebox{-.7ex}{\em $\stackrel{{\textstyle <}}{\sim}$} }
\def\leqsim{ \raisebox{-.7ex}{\em $\stackrel{{\textstyle <}}{\sim}$} }
\def\nn{\nonumber}
\def\be{\begin{equation}}
\def\ee{\end{equation}}
\def\bea{\begin{eqnarray}}
\def\eea{\end{eqnarray}}
%